\documentclass[11pt,twoside]{article}
\usepackage{atmp04}
\copyrightnotice{2004}{8}{345}{395}	
\usepackage{amscd,verbatim}
\usepackage[all]{xy}
\usepackage{graphicx}
\usepackage{amsmath}
\usepackage{amsthm}
\usepackage{euscript}

\usepackage{latexsym}

\usepackage{epsfig}
\usepackage{amssymb,amsmath,amsthm,amscd}

\usepackage{xypic}
\input{xypic}

\renewcommand{\theequation}{\arabic{section}.\arabic{equation}}
\renewcommand{\(}{\begin{equation}}
\renewcommand{\)}{end{equation} \vspace{-.05in}\linebreak}
 
\newcounter{saveeqn}
\newcounter{savealpheqn}

\newcommand{\alpheqn}{\setcounter{saveeqn}{\value{equation}}%
  \stepcounter{saveeqn}\setcounter{equation}{0}%
  \renewcommand{\theequation}{\mbox{\arabic{section}.\arabic{saveeqn}
\alph{equation}}}
  \renewcommand{\)}{\end{equation}}}
\def\part#1{\frac{\partial}{\partial{#1}}}%
\def\group#1{\refstepcounter{equation}\setcounter{saveeqn}{\value{equati 
on}}%
  \label{#1}\setcounter{equation}{0}%
 \renewcommand{\theequation}{\mbox{\arabic{section}.\arabic{saveeqn}
\alph{equation}}}
  \renewcommand{\)}{\end{equation}}}
\newcommand{\reseteqn}{\setcounter{equation}{\value{saveeqn}}%
  \renewcommand{\theequation}{\arabic{section}.\arabic{equation}}%
  \renewcommand{\)}{\end{equation}}}

\newcommand{\aalpheqn}{\setcounter{saveeqn}{\value{equation}}%
  \stepcounter{saveeqn}\setcounter{equation}{0}%
  \renewcommand{\theequation}{\mbox{
        \Alph{subsection}.\arabic{saveeqn}\alph{equation}}}
   \renewcommand{\)}{\end{equation}}}
\newcommand{\areseteqn}{\setcounter{equation}{\value{saveeqn}}%
  \renewcommand{\theequation}{\Alph{subsection}.\arabic{equation}}%
  \renewcommand{\)}{\end{equation}}}

\renewcommand{\(}{\begin{equation}}
\renewcommand{\)}{\end{equation}} \newcommand{\ba}{\begin{eqnarray}}
\newcommand{\ea}{\end{eqnarray}}

\renewcommand{\r}{\rho} \newcommand{\bp}{\mathop{\vtop{\ialign{##\crcr
   $\hfil\displaystyle{}\hfil$\crcr\noalign{\kern-13pt\nointerlineskip}
   \BIG{(}\hskip0pt\crcr\noalign{\kern3pt}}}}}
\newcommand{\cbp}{\mathop{\vtop{\ialign{##\crcr
   $\hfil\displaystyle{}\hfil$\crcr\noalign{\kern-13pt\nointerlineskip}
   \BIG{)}\hskip0pt\crcr\noalign{\kern3pt}}}}}
\newcommand{\pa}{\mathop{\vtop{\ialign{##\crcr
    
$\hfil\displaystyle{\oplus}\hfil$\crcr\noalign{\kern+1pt\nointerlineskip 
}
   \hspace{.08in}$^{\alpha=0}$\hskip6pt\crcr\noalign{\kern3pt}}}}}

\newcommand{\R}{\ensuremath{\mathbb R}}

\newcommand{\C}{\ensuremath{\mathbb C}}

\newcommand{\Q}{\ensuremath{\mathbb Q}}  
\newcommand{\Z}{\ensuremath{\mathbb Z}}

\def\r{\rightarrow}

 \newcommand{\del}{\ensuremath{\partial}}

 \newcommand{\beq}{\begin{equation}}
\newcommand{\beg}[2]{\begin{equation}\label{#1}#2\end{equation}}
\newcommand{\eeq}{\end{equation}}

\numberwithin{equation}{section}

\newcommand{\rref}[1]{(\ref{#1})}

\catcode`\@=11
\def\vereq#1#2{\lower3pt\vbox{\baselineskip1.5pt \lineskip1.5pt
\ialign{$\m@th#1\hfill##\hfil$\crcr#2\crcr\sim\crcr}}}
\catcode`\@=12

\makeatletter \newcommand\figcaption{\def\@captype{figure}\caption}
\newcommand\tabcaption{\def\@captype{table}\caption}
\makeatother
\renewcommand{\(}{\begin{equation}}
\renewcommand{\)}{\end{equation}} 

\newcommand{\CC}{{\mathbb C}}
\newcommand{\RR}{{\mathbb R}}
\newcommand{\ZZ}{{\mathbb Z}}

\theoremstyle{plain}

\theoremstyle{definition}

\begin{document}



\arxurl{hep-th/0404013}


\title{M-theory, type IIA superstrings, and elliptic cohomology}

\vspace{-.6in}

   \author{Igor Kriz}

\vspace{-.25in}


\addressemail{ikriz@umich.edu .}

\pagestyle{myheadings}
\markboth{M-THEORY AND ELLIPTIC COHOMOLOGY} {I. KRIZ AND H. SATI}

\vspace{-.25in}

\address{Department of Mathematics\\
            University of Michigan\\
            Ann Arbor, MI 48109,\\ 
            USA}
\vspace{-.25in}

\author{Hisham Sati}

\vspace{-.25in}

\addressemail{hsati@maths.adelaide.edu.au.}
\vspace{-.25in}

\address{Department of Physics\\
       University of Adelaide\\
       Adelaide, SA 5005,\\ 
       Australia}
\vspace{-.25in}

\address{Department of Pure Mathematics\\
       University of Adelaide\\
       Adelaide, SA 5005,\\ 
       Australia}



\begin{abstract}
The topological part of the M-theory partition function was shown by Witten
to be encoded in the index of an $E_8$ bundle in eleven dimensions. This
partition function is, however, not automatically anomaly-free. We observe
here that the vanishing $W_7=0$ 
of the Diaconescu-Moore-Witten
anomaly \cite{dmw} in IIA and compactified M-theory partition function is
equivalent to orientability of spacetime with respect to (complex-oriented)
elliptic cohomology. Motivated by this, we define an elliptic cohomology
correction to the IIA partition function, and propose its relationship to
interaction between $2$- and $5$-branes in the M-theory limit.
\end{abstract}

\cutpage
\newcommand{\cform}[3]{\begin{array}{c}
{\scriptstyle #3}\\
#1\\
{\scriptstyle #2}\end{array}}

%


\section{Introduction}

There are two superstring theories with two supersymmetries in ten
dimensions. One is nonchiral $N=(1,1)$ type IIA string theory and the other
is chiral 
$N=(2,0)$ IIB theory. Their fields fall into two sectors, the RR
(Ramond-Ramond) sector and the NS (Neveu-Schwarz) sector. The RR fields
naturally couple to D-branes \cite{Pol} of even (odd) spatial dimension in
type IIA (IIB) and the NS fields to the 
NS-branes. The classification of the RR fields 
has been an area of active research in the past several
years. The charges can be classified, in the absence of the NS fields, by
K-theory of 
spacetime \cite{MM}, namely by
$K^0(X)$ for type $IIB$ \cite{Wi1} and by $K^1(X)$ for type $IIA$ \cite{Ho}.
The RR fields are also classified by  K-theory \cite{MW, FH}, with the roles
of $K^0$ and $K^1$ interchanged. In the presence of the NS B-field, or its
field strength $H_3$, the relevant K-theory is 
twisted K-theory, in the sense of \cite{dk, Ros}, as was shown in \cite{FW,
Ka} by analysis of worldsheet anomalies for the case the NS field 
$[H_3] \in H^3(X, \mathbb Z)$ is a torsion class, and in \cite{BM} for the
nontorsion case. Other versions of K-theory show up as well. 
For example, KO and KSp theories are relevant for Type I \cite{Wi1},
equivariant K-theory for orbifolds, Atiyah's Real K-theory for orientifolds
\cite{guk, ho}. \footnote{A review of early developments can be found in
\cite{sz}.}

\vspace{3mm}
Eleven-dimensional M-theory \cite{m1,m2,m3}
has four BPS objects:
the membrane M2, the fivebrane 
M5, the Kaluza-Klein monopole MKK and the gravitational wave MW. 
Compactification on a circle leads to type IIA string theory. The objects in
the theory, i.e. the M-branes reduce to the IIA D-branes. 

\vspace{3mm}
D-branes and M-branes can have anomalies associated to them. This prevents
the existence of 
nonzero or 
well defined partition functions. For the M5 brane
\cite{flux, duality}, the partition 
function 
\footnote{Hopkins and Singer \cite{hs} have 
analyzed the M5 partition function, and constructed the line bundle on the
intermediate Jacobian even when the fourth integral cohomology has torsion. 
Diaconescu, Freed and Moore \cite{dfm} have also recently 
analyzed the M5 anomaly in the context of their 
model for the M-theory 3-form $C_3$.}
is nonzero if the M5 can be decoupled from the bulk, including the M2 brane.
In type IIA, the D-brane anomaly cancellation \cite{FW} is the vanishing of
the third integral Stiefel-Whitney class $W_3$. In the presence of a NS
field, it becomes $[H_3] - W_3=0$. This can be interpreted in the context of
(twisted) K-theory Atiyah-Hirzebruch spectral sequence (AHSS) as the
vanishing of the third differential $d_3$. 
This contains a topological part $W_3$
and a ``twisting'' part, the NS 3-form $H_3$.

\vspace{3mm}

On the other hand, there is another anomaly found by Diaconescu, Moore and
Witten \cite{dmw} 
in the context of relating the K-theory of ten-dimensional spacetime $X$ of
type IIA to 
partition function of M-theory on an eleven-dimensional manifold $Y$ in the
topological 
sector captured by an $E_8$ gauge theory. This anomaly is an obstruction to
the partition 
function being well-defined. The condition is the vanishing of the seventh
integral Stiefel-Whitney class $W_7$ of $X$, considered as the base of a
circle bundle with total space $Y$. In \cite{dmw} the NS fields were set to
zero for the most part. In \cite{ms} flat NS 
potentials were added, including a (cohomologically trivial) NS field
$H_3=dB_2$. However, 
the authors did not discuss the anomaly issue in that paper. Later, in
\cite{hm}, it was shown 
that the anomaly would still persist in the twisted K-theory setting.

\vspace{3mm}
A completely different motivation is to what extent K-theory can describe
D-branes, and in 
particular the fields and the charges, and whether one should seek
alternatives 
in some cases where K-theory breaks down and/or can not see the whole
picture. 
For example, on Calabi-Yau spaces there currently seems to be an
alternative: 
the derived category of coherent sheaves \cite{doug}, and in the presence of
a NS field, the 
derived category of twisted sheaves \cite{katz}. The advantage of using the
derived category 
picture of $D$-branes as opposed to K-theory, is that the former picture
contains considerably more data 
\footnote{e.g. related to geometry and $D$-brane processes.} 
that specifies the D-brane (e.g. \cite{asp}). On non-geometric backgrounds,
namely discrete-torsion D-branes there is a search for a ``quantum
K-theory''  \cite{dist}, and some other form of generalized cohomology is 
already being of relevance: $\RR/\ZZ$ generalized cohomology theory ({\it
cf.} \cite{lot}). It is natural then to suspect that on general topological
spaces an alternative of K-theory is 
needed, for example to describe phenomena at strong coupling and at the
quantum level, perhaps more ``M-theoretic'' in nature.

\vspace{3mm}
In the present paper, we study the role of a new possible tool for
investigating the $M$-theory partition function, namely elliptic cohomology.
The first motivation which led us to consider elliptic cohomology was the
question whether 
there is a twisting part that can be added to the topological part $W_7$ in 
analogy to $H_3$ in the brane-anomaly. The immediate candidate would be the
Hodge dual 
$H_7 = *_{10} H_3$ of $H_3$. One might also naively ask whether $W_7$
analogously comes from 
a differential of degree 7, $d_7$ in AHSS. But in K-theory AHSS 
of a $10$-dimensional $Spin$ manifld, $d_7$ is zero as can be seen from the 
following simple arguement.
The AHSS is double-graded, $E_{p,q}$, where
$p$ is the filtration degree (which is in the range 0 to 10, the dimension
of the manifold), and is 2-periodic in the $q$-direction. 
Thus, only even $q$-degrees occur and negative $q$-degrees occur as well.
Now a differential $d_r$ raises the $p$-degree by $r$, and decreases the
$q$-degree by $r-1$. Hence, $r$ must be odd. In any case, if the $p$-degree
is to be raised by 7, the only possible sources are 1 and 2 (the degrees 0
and 10 must be unaffected by differentials, since by K-theory valued
Poincar\'e duality these K-theory classes survive). So $d_7$ in the
cohomology K-theory AHSS can originate only in dimension 1 or 2. But in
dimension 1, every integral cohomology class is represented by a K-theory
class. This is because $H^1(M,\ZZ)$ classifies homotopy classes of maps $M\r
S^1$, and the K-theory of $S^1$ is a free module on one (odd) generator,
which we can pull back to M. But a similar argument also works in dimension
2: $H^2(M,\ZZ)$ is represented by a map $M\r \CC P^\infty$ and the K-theory
cohomology AHSS collapses for $\CC P^\infty$, in particular the generator of
$H^2(\CC P^\infty,\ZZ)$ is represented by a K-theory class, so its pullback
represents the 2-dimensional integral cohomology class in M. 
So $W_7$ can not possibly come from K-theory AHSS. However, the question
will turn out not to be that naive and we will show that it indeed comes
from a $d_7$, not in K-theory 
AHSS, but rather in Morava K-theory and elliptic cohomology! (Those are
examples of 
generalized cohomology theories, which we explain in Appendix \ref{B}.)

\vspace{3mm}
More concretely, we noticed (see Sections \ref{32}, \ref{elle} below) that
the vanishing of the Diaconescu-Moore-Witten obstruction $W_7=0$ is
precisely 
equivalent to orientability of the $Spin$-manifold $X$ with respect to 
(complex-oriented) elliptic
cohomology. This is a generalized cohomology
theory which, in addition to the Bott-periodicity element $v_1$, contains
another periodicity element $v_2$ of dimension $6$. Moreover, $v_2$ is
invertible in elliptic cohomology, while $v_1$ is not.

\vspace{3mm}
This led us to ask questions: Is there an analogue of the $M$-theory 
partition function based on elliptic cohomology instead of $K$-theory? If
so, what additional physical information does it contain? Is elliptic
cohomology, rather than $K$-theory, the right tool for describing the path
from 
$IIA$-theory to $M$-theory? 

\vspace{3mm}
We do not have complete answers to all these question, but we do obtain
clues for some of them. We do define, in Section \ref{s5} below, a lift of
the partition function of \cite{dmw} to elliptic cohomology. We focus on the
compactified case $Y=X\times S^1$, in which case \cite{dmw} show that the
$M$-theory partition function is equivalent to that of IIA. We prefer, in
fact, to work on the IIA-side of \cite{dmw}, as that is where K-theory
``lives'' \footnote{ there one has the fundamental string mode expansion to
work with, as we explain in Section \ref{eint}.}. We show that the partition
function construction in Section 7 of \cite{dmw} can be lifted to elliptic
cohomology. To be completely precise, this still requires that a
$Spin$-manifold with $W_7=0$ is orientable with respect to {\em real}
elliptic cohomology; we explain the difference in Appendix \ref{B}. As it
turns out, orientability with respect to real elliptic cohomology requires
another condition, $w_4=0$. When this condition is violated, there is an
additional anomaly. A discussion of the anomaly from the mathematical point
of view is given in Section \ref{elleo} below.

\vspace{3mm}
Now it appears that 
the elliptic cohomology-based partition function may indeed be more closely
tied to $M$-theory than the $K$-theory partition function. For one thing,
one can see directly (Section \ref{s5}) that the elliptic partition
function, when it exists, 
is not anomalous. (The distinction between existence and lack 
of anomaly may appear to be a fine one, but the point is
in the $K$-theory case, there is a condition of vanishing of a certain
quadratic function on torsion - this leads to $W_7=0$. In the elliptic
cohomology setting, $W_7=0$ is a condition required for orientability, but
when satisfied, no other condition on torsion appears.) Additionally, in
elliptic cohomology, one has $E_0=\Z[v_{1}^{3}v_2^{-1}]$, so the partition
function can be thought of as a family of partition functions indexed by a
parameter $v_{1}^{3}v_2^{-1}$. So one can ask what is the physical meaning
of these additional modes? The interpretation of $v_2$ as element in the
$6$-dimensional complex cobordism group suggests that these modes are
related to interactions between an $M2$-brane and $M5$-brane. Out of the
many possible configurations allowed by supergravity, we think the right
ones are intersections between an $M2$-brane and $M5$-brane on a string,
which, in the $S^1$-compactified case connects with the fundamental
IIA-string. Moreover, this suggests that the theory can indeed be
$H_7$-twisted, namely when the $M5$-brane is not complex-oriented. Even in
the untwisted case, however, we are seeing an anomaly of such states when
$w_4\neq 0$. We also think that the non-invertibility of the Bott element in
elliptic cohomology is related to the disappearance of some of the RR
charges when we pass from IIA to $M$-theory: these charges can be
constructed one from another in any theory in which the Bott element is
invertible. These ideas are given in Section \ref{eint} below.

\vspace{3mm}
The present paper relies very heavily on the methods of \cite{dmw}. To make
the paper self-contained, we review these methods in Sections 
\ref{sdmw}, \ref{sw3}. We also require quite a lot of information from
homotopy theory. Orientations are reviewed in Appendix \ref{A} and
additional comments salient to the present investigation are given in
Section \ref{so}. Generalized cohomology theories, complex orientations, and
formal group laws are reviewed in Appendix \ref{B}. Section \ref{elle}
contains the conclusion of the argument that a $Spin$-manifold with $W_7=0$
is orientable with respect to elliptic cohomology, and Section \ref{elleo}
describes the even much more subtle story surrounding $w_4$.

\section{The Diaconescu-Moore-Witten anomaly}
\label{sdmw}
%
To make this paper as self-contained as possible, in this section we outline
(only the relevant) results that we need from Diaconescu-Moore-Witten
\cite{dmw}. M-theory is defined on a circle bundle $Y$ with Type $IIA$ on
the base $X$. In M-theory there is an $E_8$ bundle so associated to it is
the $p_1/2$ class $\lambda$. In Type $IIA$ there are the RR fields, which
are reductions of that class, and are classified by (twisted) K-theory. The
nontorsion parts of the partition functions are compared and the anomalies 
are identified.

\subsection{The M-theory side}
First let us look at the M-theory side (compactified to $X$). Stong has
shown that 
${\tilde \Omega}_{10}^{spin} \left( K(\ZZ,4) \right)=\ZZ_2 \times \ZZ_2$.
This implies that there are two independent $\ZZ_2$-valued invariants of the
pair $(X,a)$, $a \in H^4(X, \ZZ)$ : 
\begin{enumerate}
\item $v(a)=\int_X a \cup w_6 = \int_X a \cup Sq^2 \lambda$, where the
second equality is due to X being spin. This is a linear function \( v(a +
b) = v(a) + v(b) 
\)
\item In $8k+2$ dimensions , index$D_{V_{\RR}}$ is a topological invariant 
mod 2,  name $f(a)$.
\begin{eqnarray}
f(a+b)&=&f(a) + f(b) + \int_X a \cup Sq^2 b\\
     &=&f(a) + f(b) + \int_X Sq^2 a \cup b
\end{eqnarray}
where $f(a)$ is not a linear function.
\end{enumerate}

The form  $Q(a,b)=f(a+b) - f(a) - f(b)$ is a homeomorphism from $\Omega_{10}
\left( K(\ZZ,4) \times K(\ZZ,4) \right)$ to $\ZZ_2$. $Q$ vanishes if either
$a$ or $b$ is zero. This implies that $Q$ is a homeomorphism to $\ZZ_2$ of
the relative bordism group \( \Omega_{10}^{spin} \left( K(\ZZ,4) \times
K(\ZZ,4),  K(\ZZ,4) \times 
\{*\} \times \{*\} \times K(\ZZ,4) \right)
\)
which is calculated in \cite{dmw} to be $\ZZ_2$. One of them is 
$Q(a,b)=\int_X a \cup Sq^2 b$, which is nonzero, e.g. on $X=S^2 \times S^2
\times \CC P^3$.

Now what if $a$ is a torsion class. Then $Q(a,b)$ is a torsion 
pairing 
\(
T : H_{tors}^k (X, \ZZ) \times   H_{tors}^{n-k+1} (X, \ZZ) \rightarrow
U(1).
\)

This is because $\beta(Sq^2 b)= Sq^1Sq^2b=Sq^3b$ by Adem relations. Now
$Sq^3b$ is a torsion class, and so $Q(a,b)=T(a,Sq^3 b)$. The partition
function vanishes unless $Q=T=0$. This implies the condition
$Sq^3\lambda=0$. Therefore \cite{dmw}, M-theory on a spin manifold of the
form $X \times S^1$ 
is inconsistent if $W_7 \neq 0$.

\subsection{The IIA side}
\label{22}
The relevant calculations here are K-theory calculations \cite{dmw} (and
similarly \cite{hm} for the twisted case). 
Recall that we have the mod 2 index $I(v)$ of the Dirac operator with values
in a real vector bundle $V$. For any $v \in KO(X)$, one can define the mod 2
index $I(v)$ of the Dirac operator with values in $v$. 
For any $x\in K(X)$ one has $x \otimes {\bar x} \in KO(X)$, so one can
define $j(x)=I(x \otimes {\bar x})$. Define the $\ZZ_2$-valued function
$\Omega(x)=(-1)^{j(x)}$, which satisfies \beg{eoom}{
\Omega(x+y)=\Omega(x) \Omega(y) (-1)^{\omega(x,y)} 
}
where $\omega(x,y)=I(x \otimes {\bar y})$ is an integer-valued unimodular
antisymmetric bilinear form on the lattice $\Gamma=K(X)/K(X)_{tors}$. Now if
$\Omega(x)\equiv 1$ for torsion elements of $K(X)$ then it can be regarded
as a function on $\Gamma$ and so can be used to define the line bundle and
hence its section, the RR partition function. If $\Omega(x)\not\equiv 1$ on
$K(X)_{tors}$ then the Partition function of the theory vanishes upon
summing over torsion. The story is similar in the twisted K-theory case
\cite{hm} for $\Gamma_H=K(X;H)/K(X;H)_{tors}$.

\vspace{3mm}
Then \cite{dmw} showed that the M-theory anomaly and the IIA anomaly are
related in a one-to-one
fashion:

\begin{center}
{\it $W_7(X) \neq 0$ if and only if $\Omega(x)\not\equiv 1$ on
$K(X)_{tors}$}. \end{center}

The twisted case \cite{hm} is expected to result in an analogous statement.

\section{The Stiefel-Whitney classes}
\label{sw}

\subsection{The $W_3$ story}
\label{sw3}
Freed and Witten \cite{FW} have shown that an anomaly in D-branes 
is given by:
\(
W_3 + [H_3] =0
\label{3}
\)

$W_3$ is the integral class, obtained from the second mod 2 Stiefel-Whitney
class $w_2$ via the Bockstein homomorphism. This is well-explained in
\cite{dmw} (or see the analogous 
construction for $W_7$ below).
The physical interpretation of this is the following. A D-brane cannot wrap
a submanifold 
of $X$ unless the Poincar\'e dual can be lifted to K-theory. 
The anomaly comes from the fact that when this condition is not satisfied
then we can have other branes ending on the one we are 
considering and so we cannot view it in isolation, so the 
partition function is not well-defined (or -behaved).

\vspace{3mm}
There is more than one 
possible mathematical interpretation. At the level of AHSS, as we have seen
in the introduction,
it is the third differential $d_3=W_3 + [H_3]$.   
At the level of the full twisted K-theory, one should solve the extension
problem and this 
then could
be an obstruction to a K-theory lift. 
There is another mathematical interpretation that will be relevant for our
discussion of $W_7$. For a compact oriented manifold X, this is an
obstruction to being $Spin^c$. But there is another way of looking at it, in
terms of the K-theory AHSS differential. A $Spin^c$-manifold would be
K-theory orientable, so would have a K-theory homology fundamental class.
Now one can see directly that if $W_3\neq 0$, X has no K-theory fundamental
class. Stiefel-Whitney classes are conjugates of Steenrod operations by the
Thom isomorphism. But one can also look at this in terms of Poincar\'e
duality: Let $\alpha\in H^7(X,\ZZ)$ (assuming X is 10-dimensional) be a
class such that $W_3.\alpha\neq 0$. Then 
\(
      Sq^3\alpha \neq 0.
\)
But now recall that the Milnor primitives 
\footnote{see Appendix \ref{B} for the definition and properties.} $Q_i$ are
elements in the Steenrod algebra of dimension $2^{i+1}-1$, and that \(
      Sq^3=Q_1 + {\rm decomposables}.
\)
Moreover, $Q_1$ is the primary differential $d_3$ in the K-theory AHSS. So
we see that \(
      d_3(\alpha)=u
\)
in the $K^*$-AHSS, where $u\in H^{10}(X,\ZZ)$ is the dual of the fundamental
class. Dualizing, we conclude that the fundamental class of X does not lift
to K-theory homology, so X is not K-theory orientable.

\subsection{The $W_7$ story} 
\vspace{3mm}
\label{32}

This section contains some observations regarding the obstruction $W_7(X)$,
constructed in \cite{dmw} for a compact $Spin$-$10$-manifold $X$, to the
existence of a consistent theory of IIA RR $D$-branes in $X$, as well as to
having a non-zero partition function of $M$-theory on $X\times S^1$.

\vspace{3mm}
We begin by recalling a subtlety regarding the definition of $W_7$: this is
not the $7$-th Stiefel-Whitney class $w_7\in H^7(X,\Z/2)$, which always
vanishes for a $Spin$ manifold. Rather, $W_7\in H^7(X,\Z)$ is the canonical
integral lift of $w_7$, namely 

\(
\beta(w_6)
\)
where $\beta:H^k(X,\Z/2)\r H^{k+1}(X,\Z)$ is the Bockstein homomorphism. The
fact that the mod $2$ reduction $w_7$ of $W_7$ vanishes signifies the fact
that $W_7$ is divisible by $2$ in integral cohomology.

\vspace{3mm}
This is strikingly analogous to another situation: namely, for a
$Spin$-manifold $X$, the first Pontrjagin class $p_1$ is divisible by $2$,
and $\lambda=p_1/2$ is an obstruction to what \cite{kil,st} call  ``string
structure''. One may ask if there is any connection between these two
obstructions. One connection we can see right away. A string structure is
the same thing as lifting the structure group of the tangent bundle of $X$
to the $3$-connected cover 
$String(10)$ of
$Spin(10)$. This is, however, by Bott periodicity the same thing as the
$6$-connected cover, as there are no homotopy groups in between. In other
words, the classifying map $X\r BSpin(10)$ lifts to $BString(10)$, which is
$7$-connected, and therefore has no cohomology in dimension $7$. Therefore,
$W_7(X)=0$. Even more directly, one has $w_6=Sq^2\lambda$, so $\lambda=0$
implies $w_6=0$ which implies $W_7=0$.

\vspace{3mm}
One may also ask if conversely, $W_7(X)=0$ implies $p_1(X)/2=0$. However, a
moment's reflection shows that this is false. For example, for $X=S^2\times
S^2\times \C P^3$, $p_1/2$ is non-zero (and non-torsion), while there is no
odd cohomology, so $W_7=0$. 

\vspace{3mm}
We therefore further ask what is the geometric meaning of $W_7\neq 0$. To
this end, recall the definition of Stiefel-Whitney classes: If we denote by 
$D:H^k(X)\r H_{10-k}(X)$ Poincar\'e duality, we have
\(
D(w_k)=Sq^{k}(\mu)
\)
where $\mu\in H_{10}(X,\Z/2)$ is the fundamental class. Here recall the
action of Steenrod operations on homology $Sq^{k}:H_{m}(X,\Z/2)\r
H_{m-k}(X,\Z/2)$. We therefore have \(
D(W_7)=\beta_* Sq^6(\mu)
\)
where $\beta_*:H_m(X,\Z/2)\r H_{m-1}(X,\Z)$ is the Bockstein.

\vspace{3mm}
Now there is a distinguished integral cohomological operation \(
\tilde{Q}_2:H^m(X,\Z)\r H^{m+7}(X,\Z)
\)
(dually also a homological operation lowering dimension by $7$) which is 
the integral lift of the Milnor primitive $Q_2$ (see \cite{bp}, \cite{mil}).
Moreover, the operation $\tilde{Q}_2$ is closely tied to a generalized
cohomology theory $\tilde{K}(2)$ known as {\em $p=2$ integral second Morava
$K$-theory} which is a reduction of elliptic cohomology (see \cite{rav}
section 4.2, or Appendix 
\ref{B} for background). One has
\(
\tilde{K}(2)^{*}(*)=\Z[v_2,v_{2}^{-1}]
\)
where $v_2$ is in dimension $6$. The way this theory is tied to
$\tilde{Q}_2$ is as follows: as for every generalized cohomology theory,
there is a corresponding generalized homology theory, and Atiyah-Hirzebruch
spectral sequences both in homology and cohomology. Working, for example, in
homology, the AHSS for $\tilde{K}(2)$ is \(
E^{2}_{pq}=H_p(X,\tilde{K}(2)_{q}(*))\Rightarrow \tilde{K}(2)_{p+q}(X).
\)
The dimensions imply that possible differentials of this AHSS are
$d^{6k+1}$. The connection with $\tilde{Q}_2$ is that \( d^{7}=\tilde{Q}_2.
\)

\vspace{3mm}
Now, moreover, $\tilde{Q}_2$ coincides with $\beta Sq^6$ modulo elements of
lower Cartan-Serre filtration 
(see \cite{mil}). Further, however, 
note that there are only three linearly independent Steenrod operations in 
dimension $6$, namely 
\footnote{By Adem relations, one has
$Sq^2Sq^4=Sq^6+Sq^5Sq^1$ and $Sq^1Sq^5=0$.} 
$Sq^6$, $Sq^5 Sq^1$ and $Sq^4 Sq^2$. Further, on
the fundamental class $\mu\in H_{10}(X,\Z)$, the last two must vanish by the
assumption that $X$ is a Spin manifold. 
We conclude that 
\( \tilde{Q}_2(\mu)=\beta Sq^6(\mu)=D(W_7).
\)
Therefore, we see that $W_7\neq 0$ if and only if the primary differential
in the homology 
$\tilde{K}(2)$-AHSS for $X$ is non-zero on $\mu$, which in turn happens if
and only if $X$ is not 
$\tilde{K}(2)$-orientable (as any higher differentials are out of filtration
degree range). We have therefore proved 
\begin{center} 
{\it A $10$-manifold $X$ is orientable with respect to $\tilde{K}(2)$ if and
only if $W_7(X)=0$}. 
\end{center} \vspace{3mm} 

\subsection{Orientation}
\label{so}

We have seen the two anomalies in ten dimensions, namely, $W_3=0$ the one
coming from D-brane worldvolume, and $W_7=0$ the one coming from the
M-theory partition function. 
Since orientation seems to be somewhat of a unifying theme for both
anomalies, we will give some arguments on the relevance and possible
consequences of having an orientation. In Appendix \ref{A}, we have outlined
some main points regarding orientations in any (generalized)
cohomology. First let us consider the simplest case which is just the
``usual'' orientation   
of the manifold. One instance (aside from torsion issues) where orientation
plays a role is in supersymmetry. For example, eleven-dimensional
supergravity on $AdS_4 \times N^7$ where $N^7$ is an Einstein manifold with
killing spinors, can have different amount of supersymmetry depending on the
orientation of   
$N^7$, and in some cases one of the two orientations leads to no
supersymmetry at all.

\vspace{3mm}
One would like to be able to relate the cohomology of the branes to that of
spacetime and vice versa. The Thom-Dold \footnote{Dold generalized the Thom
isomorphism and the Chern classes to generalized cohomology.} isomorphism
and the Gysin homomorphism provide that link, without which it would be hard
to study phenomena like anomaly inflow. A consequence of this is also the
existence of Poincar\'e duality. This is desirable in order to be able to go
from the description of D-branes as homology cycles and the description of
fields and charges in term of cohomology. The formulae for the charges
depends on characteristic classes, which are most elegant in the case when
all the vector bundles are $E$-orientable. One consequence of viewing the
problem as that of orientability is that certain properties follow
immediately. For example, applying the theorem that says that any manifold
is $E$-orientable if and only if its stable normal bundle is $E$-orientable,
to the case $E=K$, 
one sees that
the normal bundle being $spin^c$ is a direct consequence of the manifold
being so, and vice versa.

\vspace{3mm}
One can then define integers out of the characteristic classes, by pairing
them with the fundamental class as in Appendix \ref{A}. Of course, this is
needed, e.g., to get integral charges for the branes. Another consequence is
that one can use different orientations to get those charges. It seems
reasonable then to believe that given an $E$-orientation 
of a manifold then one can find some integral formula, and changing the
orietation to some $E'$ then gives the charges in some other form, so we
expect that the formula for the charges can be derived using Morava K-theory
orientation and/or elliptic cohomology orientation. One might further
suspect that the higher generalized cohomology theories could perhaps as
well give corrections to the charge formula derived using K-theory.

\vspace{3mm}
Since the Steenrod square $Sq$ is a ring morphism, one can apply
(\ref{integral}) to this case, \( \langle x, [X] \rangle=\langle
Sq(x)W(\nu),[X] \rangle,
\)
which when $\dim x \neq n$, i.e. the case relevant for the M-theory 4-form,
reduces to \footnote{All relations among those characteristic classes follow
from this \cite{BP2}.} \( \langle Sq(x)W(\nu),[X] \rangle=0.
\)
Now when $Sq^1(x)=0=Sq^2(x)$, the condition becomes $Sq^3(x)W^3(\nu)=0$.
This implies that either $x$ has a K-theory lift \cite{dmw} or that
$W^3(\nu)$ vanishes, (or both of course).

\vspace{3mm}
One can relate orientations with respect to one cohomology theory to
orienations with respect to another. For example, from the ring morphism
$KO\r K$, any $KO$-orientable object is also $K$-orientable. This can be
translated to the familiar language: every $spin$ manifold is also $spin^c$.

\subsection{Realizability} 
\label{34}

The class $W_3$ can be realized by vector bundles, namely by the universal
oriented
vector bundle. If $W_3(\xi)=0$ for some $H\ZZ$-oriented vector bundle $\xi$,
then   
$\xi$ is $K$-orientable. 
One possible consequence of this is the question of interpreting anomalies
as obstruction to having a section on (determinant) bundles. In \cite{FW}
the $W_3$ anomaly was interpreted using the line bundle ${\rm
Pfaff}(D)\otimes {\mathcal{L}}_B$ associated to the Dirac operator $D$ and
the $B$-field. This suggests that the Diaconescu-Moore-Witten  
anomaly in M-theory cannot be represented in such a formalism. 
This hints that in the context of the $W_7$ anomaly, instead of
(virtual) vector bundles, we should perhaps deal directly with classes in a
generalized cohomology theory.

\section{Elliptic cohomology}

\label{s5}

It is natural to conjecture that in the result of Section \ref{32},
$\tilde{K}(2)$ can be replaced by elliptic cohomology $E$. This is in fact
true, as can be proved by more technical manipulation of the AHSS (see
Section \ref{elle}). The first question one has to settle 
is which definition
of $E$ we should choose. However, our characteristic classes observations
give some clues. In particular, note that $W_7=0$ for any compact complex
$10$-manifold, so we should use {\em complex oriented elliptic cohomology}.
On the other hand, recall that the Hopkins-Miller universal elliptic
cohomology theory $tmf$ \cite{Hm} is $MO\langle 8\rangle$-orientable, which
means that every manifold whose stable normal bundle has structure group
$String=O\langle 8\rangle$ is $tmf$-orientable. Thus, we have an explanation
of the above distinction between the $p_1/2$ and $W_7$ obstructions: it
signifies that our current level of observation sees a connection between
$M$-theory and complex-oriented elliptic cohomology, but not (yet) $tmf$.

\vspace{3mm}
Now there are still various models for complex-oriented cohomology $E$ which
are characterized by their coefficient rings. Under certain basic
assumptions (essentially, the associated elliptic curve must allow both
multiplicative and $p=2$-supersingular reduction), 
these theories contain equivalent homotopical
information, but each has some advantages or disadvantages. Choosing a
complex-oriented elliptic cohomology theory is akin to choosing coordinates.
For a fuller discussion, we refer the reader to Appendix \ref{B}. The
problem is that there cannot be a {\em univeral} complex-oriented elliptic
cohomology theory. This is explained quite well in \cite{tmf}. Although
there is in some sense a universal (generalized) elliptic curve, called the
Weierstrass curve given by the equation $$y^2x+a_1 xyz +a_3 y^3=x^3+a_2
x^2z+a_4 xz^2 +a_6 z^3$$ over the ring $\Z[a_1,a_2,a_3,a_4,a_6]$, and we may
choose accordingly\\ $E_*=\Z[a_1,a_2,a_3,a_4,a_6][u,u^{-1}]$, this theory
still cannot be considered universal because of automorphisms. However, the
parameters $a_i$ are
(generalized) modular forms, and there is a ``character map''
\begin{equation} \label{char} E\r K[[q]][q^{-1}] \end{equation} where $K$ is
$K$-theory, $q$ is a parameter of dimension $0$, $K[[q]]$ is therefore a
product of infinitely many copies of $K$, and 
the notation $[q^{-1}]$
signifies that $q$ is inverted. The map (\ref{char}) is determined by what
happens on coefficients, which can be found for example in \cite{tmf},
Section 2.6, see also Appendix \ref{B}.

\vspace{3mm}
The Weierstrass equation (and the definition of $E$) can be simplified 
substantially if the prime $6$ is inverted. Unfortunately, this is not
suitable for us, as $p=2$ information is critical to to our investigation
(as seen already in the $K$-theory calculation of the IIA partition function
described in \cite{dmw}). Another kind of simplification arises if we are
willing to {\em complete} at the prime $2$. For example, we may then choose
$E_*=W_2[[a]][u,u^{-1}]$ where $W_2$ is the ring of Witt vectors, i.e. the
ring of integers of the extension of the field $\Q_2$ of $2$-adic numbers by
an Eisenstein polynomial of degree $4$, $dim(a)=0$, $dim(u)=2$. The
characteristic property of this theory is that its formal group law,
calculating $c_1(\xi\otimes\eta)$ from $c_1(\xi)$, $c_1(\eta)$ for line
bundles $\xi,\eta$ is a {\em Lubin-Tate law} $F_2$ of height $2$ (see
\cite{lt}). One can construct a character map (\ref{char}) for this theory
with the only exception that $K$ must be replaced by $K$-theory with
coefficients in $W_2$ (see again Appendix \ref{B} for more details). 

\vspace{3mm}
An even further simplification may be obtained if we notice that 
the cohomology theory $E$ of
the last paragraph (with $E_*=W_2[[a]][u,u^{-1}]$) is a completion of a
finite sum of suspensions of copies of the cohomology theory $E(2)$ which
has $E(2)_*=\Z[v_1,v_2,v_2^{-1}]$ where $v_1$ has dimension $2$ and $v_2$
has dimension $6$. One may set, for example, $v_2=u^3$, $v_1=au$. This
cohomology theory lacks some of the manifest modular symmetries of the other
elliptic cohomology theories, (in particular is only $6$-periodic in
dimension), but has the simplest coefficients and we will often find it most
convenient for our purposes.

\vspace{3mm}
One more refinement is relevant. Throughout our discussion, we are dealing
with $Spin$ manifolds, and hence some real structure is expected to 
play a role. Therefore, we need to consider a {\em real form} $EO(2)$ of
elliptic cohomology, which is obtained, roughly speaking, by taking the
fixed point cohomology theory of $E$ under $\Z/2$-action which comes from
the FGL isomorphism $-i(x)$ where $i(x)$ is the inverse series of $F_2$. In
fact, making such definition precise is difficult, as one must prove the
$\Z/2$-action is rigid, and not just up to homotopy. However, a rigorous
definition of $EO(2)$ is given in \cite{hk}, and its coefficients $EO(2)_*$
are calculated there also. 
To construct our elliptic partition function, {\em from now
on, we shall assume that $X$ is orientable under $EO(2)$.} One could
conjecture this is true if and only if $W_7=0$ and $X$ is $Spin$, but it
turns out that the situation is not quite as simple as that. Instead, a
$Spin$ manifold is $EO(2)$-orientable if and only if it satisfies $w_4=0$
(we prove this in Section \ref{elleo} below). In other words, when $w_4\neq
0$, we see yet another anomaly to the existence of an elliptic cohomology
partition function.

\vspace{3mm}
Now recall from \cite{dmw} and Section \ref{22} above that the key point of
the construction of the RR paritition function of IIA $D$-branes is the
function $\Omega$ (or equivalently $j$) defined on $K^0(X)$.
Homotopy-theoretically, the construction of $j$ means that for $x\in
K^0(X)$, the virtual bundle $x\otimes \overline{x}$ has a real structure,
and hence represents an element of $KO^0(X)$. The mod $2$ index is simply
the Kronecker product with the 
$KO$-orientation $KO$-homology class $\mu\in KO_{10}(X)$:
$$KO^0(X)\otimes_{KO_{*}} KO_{10}(X)\r KO_{10}(*)=\Z/2.$$ The construction
of the partition function fails when this index is non-zero on torsion
elements $x\in K^0(X)$.

\vspace{3mm}
Let us now mimic the construction of the theta-function described in
\cite{dmw}, Section 7 in elliptic cohomology. To this end, as suggested in
Section \ref{34} above, we shall deal directly with generalized cohomology
classes, in this case with $E^0(X)$-classes. The manifold $X$ is
$E$-orientable, so it has an orientation class $[X]_E\in E_{10}X$. For
$x,y\in E^0(X)$, we put 
$$\omega(x,y)=\langle x \overline{y},[X]_E\rangle\in E_{10}=E_0$$ Now we
need an elliptic refinement of the function $j$. Assuming 
that $X$ is $EO(2)$-orientable, we have an $EO(2)$-orientation class
$[X]_{EO(2)}\in EO(2)_{10}(X)$. Now for $x\in E^0(X)$, the class
$x\overline{x}$ lifts canonically to $EO(2)^{0}(X)$, so we may put
$$j(x)=\langle x\overline{x},[X]_{EO(2)}\rangle\in EO(2)_{10}.$$ To see what
the right hand side is, we must compute $EO(2)_{10}$. 
This was done in
\cite{hk}. It is helpful for the purpose of the computation to reduce $E$ to
a theory $E(2)$ with coefficients 
$\Z_{2}[v_1,v_2,v_{2}^{-1}]$ (it is a direct summand of $E$, so this will
allow us to omit repeating terms). Now one more relevant piece of
information is the {\em twist}. The correct way of viewing the real version
of the theory $E(2)$ is as a $\Z/2$-equivariant generalized cohomology
theory, which we denote by $E\R(2)$. Cohomology classes of such theory are
indexed by $k+\ell\alpha$ where $\alpha$ denotes the sign representation of
$\Z/2$ (see \cite{hk}). Then the orientation class of $X$, which we assume,
is in $E\R(2)_{10}(X)$, so the Kronecker product lies in $E\R(2)_{10}(*)$.
In the notation of \cite{hk}, this group is a $\Z/2$-vector space with basis
\begin{equation} \label{generators1} v_{1}^{3n}v_{2}^{2-n}\sigma^{-4}a^2, \;
n\geq 1. 
\end{equation}
Here $v_1$ has dimension $1+\alpha$, $v_2$ has dimension $3(1+\alpha)$,
$\sigma$ has dimension $\alpha-1$ and $a$ has dimension $-\alpha$, so the
generators (\ref{generators1}) have dimension $10$. See Appendix \ref{B} for
more on the theory $E\mathbb{R}(2)$. In any case, we have produced an
element 
$j(x)\in \Z/2[v_{1}^{3}v_{2}^{-1}]=EO(2)_{10}$.

\vspace{3mm}
To be able to say that this element $j$ is an elliptic refinement of the
element constructed in \cite{dmw}, i.e. to produce, upon a choice of the
parameter $v_2$, a function $\Omega(x)$ with \rref{eoom}, we must make sense
of the identity
\beg{eoom1}{j(x) +j(y) -j(x+y) \equiv \omega(x,y) \mod 2.}
Note that this is more delicate than in the $K$-theory case, as we do not
(at least at present) have index-theoretical arguments at our disposal for
$E(2)$, and the two sides of \rref{eoom1} apparently belong to different
generalized cohomology groups. A purely homotopy-theoretical argument,
however, is possible, and in fact its analogue is interesting in the case of
$K$-theory, too. We see that the left hand side of \rref{eoom1} is $$\langle
x\overline{y} +\overline{x}y,[X]_{EO(2)}\rangle,$$
while the right hand side is
$$\langle x\overline{y},[X]_{E(2)}\rangle.$$
It will therefore suffice if we can make sense, for any $a\in E(2)^0(X)$, of
the identity \beg{eoom2}{\langle
a+\overline{a},[X]_{EO(2)}\rangle\equiv\langle a,[X]_{E(2)}\rangle \mod 2.}
To interpret \rref{eoom2}, we consider the {\em transfer} map $$\tau:E(2)\r
EO(2).$$ In the notation of \cite{hk}, this can be interpreted for example
as a map on fixed points of the $\Z/2$-equivariant generalized cohomology
theories \beg{eoom3}{{ \diagram E\R(2)\wedge (\Z/2_+)\rto & E\R(2)\wedge
(E\Z/2_{+})\rto^{N}& F(E\Z/2_{+},E\R(2)). \enddiagram}} Here $E\Z/2$ is a
contractible space with free $\Z/2$-action. The second arrow $N$ of
\rref{eoom3} is the norm map from a Borel homology to Borel cohomology
theory (see \cite{hk}). Now for $a\in E(2)^0X$, we have the following
commutative diagram: $$ \diagram S^{10}\dto_{[X]_{EO(2)}}&&\\ EO(2)\wedge
X\dto\rto^{Id\wedge a} & EO(2)\wedge E\dto\rto^{Id\wedge\tau}& EO(2)\wedge
EO(2) \dto\\ E\wedge X\rto^{E\wedge a} & E\rto^{\tau} & EO(2). \enddiagram
$$ The unlabelled arrows are the forgetful map and multiplications. Now this
diagram shows that \rref{eoom2} is valid if we map the right hand side to
the left hand side using the transfer! But is the transfer non-trivial on
the elements concerned? This is where \rref{eoom3} comes in. It is shown in
\cite{hk} that the norm map can be calculated by dividing by $\sigma a$ the
differential in the Tate cohomology spectral sequence for $\widehat{E\R}$
which crosses the line between Borel homology and cohomology. The relevant
differential is $$d:\sigma^{-2}\mapsto v_{1}a^{3},$$ so we get for $n\geq
1$,
$$v_{1}^{3n}v_{2}^{2-n}\sigma^{-4}a^2=\tau(v_{1}^{3n-1}v_{2}^{2-n}\sigma^{-5
}).$$
On the right hand side, we are in $E(2)_{10}$, so the $\sigma$'s can be
dropped. We see these are precisely the generators of $E(2)_{10}$.

\vspace{3mm}
Thus, we have produced an analogue of the setup of \cite{dmw}, Section 7.1,
but over the ring $\Z[v_{1}^{3}v_{2}^{-1}]$ instead of $\Z$. We would like
to interpret this as a family of theta-functions parametrized by
$v_{1}^{3}v_{2}^{-1}$. The question is, can this parametric theta-function
be anomalous and how is it related to the theta-function of \cite{dmw},
since we are using a different generalized cohomology theory?

\vspace{3mm}
To answer the first question, it turns out that this theta-function cannot
be anomalous, although it is possible that it could be {\em twisted}, see
Section \ref{eint} below. To show that no anomaly is possible, note that
anomaly of the construction \cite{dmw}, Section 7.1 can arise when the
function $j$ is non-zero on torsion elements. We have seen that this cannot
happen when $X$ is $E$-orientable (see also a more direct argument below).
However, this is not quite enough, as a priori additional anomaly could be
related to torsion in the element $v_{1}^{3}v_{2}^{-1}$. Fortunately, this
does not happen: we will see in Section \ref{elle} below from AHSS arguments
that \begin{equation} \label{ecompp}
E^{0}(X)=k^{0}(X)[v_{1}^{3}v_{2}^{-1}]=K^0(X)[v_{1}^{3}v_{2}^{-1}],
\end{equation}
so there is no torsion in the $v_{1}^{3}v_{2}^{-1}$ element.

\vspace{3mm}
To address the second question (comparison of theta-functions), 
we invoke the character map (\ref{char}).
By (\ref{ecompp}), we see that the theta-function we construct must be but a
$1$-parametric family of ordinary elliptic functions in the parameter $q$,
by the comparison map from elliptic cohomology to $K$-theory. Discarding the
additional parameter, therefore, we obtain the partition function of
\cite{dmw}.

\vspace{3mm}
To give a direct argument for vanishing of the $K$-anomaly in the
$E$-partition function in terms of bundles, consider again a complex bundle
with real structure $x\otimes\overline{x}$. Now this cannot be identified
with an element of $EO(2)^*(X)$. On the other hand, bundles with real
structure have {\em characteristic classes} in 
$EO(2)^*(X)$. Which characteristic class should we choose? There is such
characteristic class for every polynomial in Chern classes. The description
of that class we choose depends in part on our complex orientation, i.e. on
the formal group law on $E_*$: let us therefore assume, without loss of
generality, that the law, upon reduction to Morava $K(2)$-theory, has
\begin{equation} \label{FGL}x+_{F_2}x=2x + u^{3}x^4. \end{equation} (This
means that the elliptic cohomology has an {\em automorphism} which turns the
Lubin-Tate law into the FGL, which after reduction to Morava $K(2)$-theory,
is (\ref{FGL}).) Now Chern classes are, in any complex-oriented cohomology,
determined by their values on direct sums of line bundles. The class we are
interested in is the {\em determinant class} $D$ given by \begin{equation}
\label{det} D(L_1\oplus...\oplus
L_n)=c_1(L_1)+_{F_2}...+_{F_2}c_1(L_n).\end{equation}
As remarked above, for a complex bundle with real structure we automatically
get a corresponding $EO(2)^*(X)$-valued characteristic class. Now since we
are assuming $X$ is $EO(2)$-oriented, let $[X]_{EO(2)}\in 
EO(2)_{10}X$
again
be the orientation class. We may therefore go ahead and define $j_E(x)\in
EO(2)_*$ by the formula $$j_E(x)=\langle D(x\otimes
\overline{x}),[X]_{EO(2)}\rangle.$$
The map (\ref{char}) shows that the reduction of $j_E$ to $K$-theory is the
invariant $j$ considered above. 

\vspace{3mm}
It is appropriate here to make note that we have introduced an
$\alpha$-twist. The determinant characteristic class, 
which
we selected, now lands in $E\R(2)^{1+\alpha}(X)$ where $\alpha$ denotes the
sign representation of $\Z/2$ (see \cite{hk}). Then the orientation class of
$X$, which we assume, is in $E\R(2)_{10}(X)$, so the Kronecker product lies
in $E\R(2)_{9-\alpha}(*)$. In the notation of \cite{hk}, this group is a
$\Z/2$-vector space with basis \begin{equation} \label{generators}
v_{1}^{3n-1}v_{2}^{2-n}\sigma^{-4}a^2. 
\end{equation}
Here $v_1$ has dimension $1+\alpha$, $v_2$ has dimension $3(1+\alpha)$,
$\sigma$ has dimension $\alpha-1$ and $a$ has dimension $-\alpha$, so the
generators (\ref{generators}) have dimension $9-\alpha$.

\vspace{3mm}
Now it can be seen directly that the invariant $j_E$
vanishes on torsion elements of $K^0(X)$. To see this, consider the formula
(\ref{FGL}) above. Reducing modulo $2$, the first summand vanishes. Reducing
further to Morava $K$-theory \cite{mor}, taking $2^n$-th powers is an
automorphism of the theory, and $u$ is invertible. Consequently, by
completeness, the same conclusion must hold for $EO(2)$: $j_E(2x)$ for a
bundle $x$ is just an image of $j_E(x)$ under an automorphism of the theory,
hence cannot vanish if $j_E(x)$ does not vanish. However, if $2^k x=0$ (the
only interesting torsion is at the prime $2$), then $j_E(2^k x)$ must
vanish, hence so does $j_E(x)$!

\vspace{3mm}
It is curious that the lifting of bundles to elliptic cohomology which
showed the vanishing of $j_E$ on twisting is non-linear, and instead has a
degree $4$ correction, coming from (\ref{FGL}). We saw above that the right
approach is to work in elliptic cohomology directly, but it would be nice to
have a direct physical interpretation of this non-linear behavior on
bundles.

\vspace{3mm}
>From the point of view of $M$-theory, it is interesting
to note that $p_{1}/2$ is the obstrution for the $LK(\Z,3)$- (hence
$LE(8)$-) bundle on $X$ induced from the tangent bundle on $Y=X\times S^1$
to lift to a $\tilde{L}E(8)$-bundle where $\tilde{L}E(8)$ denotes the
universal central extension (= the ``affine group"). If such bundle existed 
\footnote{This was considered in \cite{hm}.},
then again 
$X$ is orientable
with respect to $tmf$. From the above discussion, we see
once again that orientability with respect to $EO(2)$ 
is the more salient notion at the current level of observation.

\vspace{3mm}
More concretely, recall again that the phase
factor $f(a)$ of $a\in H^4(X,\Z)$ can be calculated by taking 
mod $2$ index of the adjoint
$e_8$-bundle $\xi$ associated with $a$. This can again be written by
considering $\xi\in KO^{0}(X)$; the mod $2$ index is then the 
Kronecker product with the fundamental class $\mu\in KO_{10}(X)$. Therefore,
when $X$ is $EO(2)$-orientable, we can again refine this invariant to
elliptic cohomology, i.e. put $$f_{E}(a)=\langle x,[X]_{EO(2)}\rangle\in
EO(2)_*$$ where $x$ is a lift of $\xi$ to elliptic cohomology. It can
therefore be expected that the entire elliptic theta-function can be seen 
also from the $M$-theoretical point of view, although a careful comparison
between both sides is needed (see Section \ref{eint} for more on this).

\vspace{3mm}
This now points to a connection with the speculations \cite{segal} that
elliptic cohomology index can be calculated as some version of index of the
Dirac operator on the associated loop bundle. Therefore, instead of an
$E_8$-bundle, we are compelled to consider the associated $LE_8$-bundle.
That bundle resides on $LX$, but \cite{segal} gives evidence that the index
information may be concentrated around constant loops. In that case, an
$LE_8$-bundle on $X$, again, corresponds to an $E_8$-bundle on $X\times
S^1$. The line of thought of \cite{segal} is that the $tmf$-index of $\xi$
(for example the Witten genus) would be calculable in this geometric way if
${p_1/2}(X)=0$. The evidence from the $M$-theory partition function which we
have gathered points to the conclusion that the corresponding $EO(2)$-index
should have such geometric interpretation if and only if $X$ is $Spin$ and
$w_4(X)=0$. It is also worth noting that a direct connection of bundles on
loop space with elliptic cohomology, via $2$-vector bundles and $K$-theory
of $ku$, was proposed in \cite{bdr}.

\section{Orientability with respect to elliptic cohomology}

\subsection{The $E$ story}
\label{elle}
In this section, we prove that a $Spin$-manifold $X$ with $W_7=0$ is
orientable with respect to complex-oriented elliptic cohomology $E$. As
explained above, instead of $E$, we may work with $E(2)$ which has
coefficients $E(2)_*=\Z_2[v_1, v_2,v_{2}^{-1}]$ where $dim(v_1)=2$,
$dim(v_2)=6$. 

\vspace{3mm}
To start the proof, first note that we already know that $X$ is orientable
with respect to $K$-theory. We next claim that $X$ is also orientable with
respect to {\em connective} $k$-theory. To see this, consider the $k$-theory
homology AHSS \begin{equation} \label{ahss1}
E^{2}_{pq}=H_p(X,k_q(*))\Rightarrow k_{p+q}X. \end{equation} We have
$k_*(*)=\Z[v_1]$. The proof uses the observation that, in fact for every
space $X$, in the spectral sequence (\ref{ahss1}), we have \begin{equation}
\label{ahss2} \text{$v_1:E^{r}_{*,q}\r E^{r}_{*,q+2}$ is onto for all $q$
and iso for $q\geq r-2$.} \end{equation} This is proved by induction on $r$.
It is true for $r=2$. So assume (\ref{ahss2}) 
is true
for a particular $r$, we try to prove it with $r$ replaced by $r+1$. Thus,
consider \( d^r:E^{r}_{p,q}\r E^{r}_{p-r,q+r-1}.
\)
First, if $d^r(v_1 x)=v_1 d^r(x)=0$, then $d^r(x)=0$ by the induction
hypothesis, so $Ker(d^r)$ is generated in $q$-degree $0$, and hence so is
$H(d^r)=E^{r+1}$, proving the ``onto'' part of the statement with $r$
replaced by $r+1$. To prove the ``iso'' part, consider $x\in E^{r}_{*,\geq
r}$ with $v_1 x=d^r y$. Then $y=v_1 z$ for some $z$ by the induction
hypothesis. So $v_1(x-d^r z)=0$, but then $x-d^r z=0$ by the induction
hypothesis, concluding the proof of the ``iso'' part. 

But now from (\ref{ahss2}), it follows that whenever $d^r(x)\neq 0$ in the
spectral sequence (\ref{ahss1}), it is also non-zero in the spectral
sequence obtained from (\ref{ahss1}) by inverting $v_1$, which is the
$K$-theory homology AHSS. In particular, since $X$ is $K$-orientable, its
homology fundamental class supports no differential in the $K$-theory
homology AHSS, hence by what we just said supports no differential in the
$k$-theory AHSS, hence $X$ is $k$-orientable.

\vspace{3mm}
Now assume as we did that $X$ has, in addition, $W_7=0$. As observed above,
then $X$ is $\tilde{K}(2)$-orientable where $\tilde{K}(2)$ is integral
Morava $K$-theory with $\tilde{K}(2)_*=\Z[v_2,v_2]^{-1}$. 
This was simply because by sparsity of dimensions,
the only possible differential on the fundamental
class was $d_7$ which was identified with the class $W_7$. Hence, the exact
same argument also proves that $X$ is orientable with respect to the {\em
connective} integral Morava $K$-theory $\tilde{k}(2)$, which has
$\tilde{k}(2)_*=\Z[v_2]$. In fact, we claim that we can prove a stronger
statement, namely that 
\begin{equation}
\label{ahss3}
\text{The $\tilde{k}(2)$-homology AHSS for $X$ collapses to $E^2$.}
\end{equation} To prove this, note that since we already know $X$ is
$\tilde{k}(2)$-orientable, the $\tilde{k}(2)$-homology and cohomology AHSS's
are isomorphic (by capping with the fundamental class, which is a permanent
cycle). So we can work in the cohomology AHSS instead. Now by sparsity of
dimensions, the first possible differential is $d_7$. Since $d_7$ cannot
disturb filtration degrees $0$ and $10$ (by orientability), the only
possible differentials are $d_7:E_{2}^{1,*}\r E_{2}^{8,*}$ or
$d_7:E_{2}^{2,*}\r E_{2}^{9,*}$. But now first and second integral ordinary
cohomology classes are represented by maps into $S^1$ and $\C P^{\infty}$
respectively, for which the $\tilde{k}(2)$-cohomology AHSS collapse.
Therefore, by naturality of spectral sequances, the $d^7$'s considered must
also be $0$, concluding the proof of ({\ref{ahss3}).

\vspace{3mm}
Now we shall consider the second Johnson-Wilson generalized cohomology
theory $BP\langle 2\rangle$ (see \cite{rav}, Section 4.2) with coefficients
$BP\langle 2\rangle_*=\Z[v_1,v_2]$. We have a cofibration sequence of
generalized cohomology theories (more precisely {\em spectra})
\begin{equation} \label{cofib} v_1\cdot BP\langle 2\rangle\r BP\langle
2\rangle\r \tilde{k}(2). \end{equation} This leads to an exact couple, and a
corresponding spectral sequence \begin{equation} \label{ahss4}
E^{2}_{p,q}=\tilde{k}(2)_{p}X\otimes k_{q}\Rightarrow BP\langle
2\rangle_{p+q}X. \end{equation} (Recall that $k_{*}=\Z[v_1]$ where $v_1$ is
in dimension $2$.) Also, we have just proved that \(
\tilde{k}(2)_*X=H_*(X,\Z)\otimes \tilde{k}(2)_*.
\)
Now we would like to claim that
\begin{equation}
\label{ahss5}
\text{The fundamental class $\mu\in \tilde{k}(2)_{10}X$ is a permanent cycle
in (\ref{ahss4}).} \end{equation} To this end, first note that there is a
comparison map from the cofibration
(\ref{cofib}) to the cofibration sequence
\(
v_1k\r k\r H\Z,
\)
and hence a comparison map from (\ref{ahss4}) to (\ref{ahss1}). This means
that any target of a differential on $\mu$ in (\ref{ahss4}) must be of the
form $v_2 a$ (as otherwise the comparison map would give a differential in
(\ref{ahss1}), which has no differential originating in $\mu$). But now
dimensional considerations show that the only possibility is \(
d_3(\mu)=v_1v_2a
\)
for some $a\in H_1(X,\Z)$. Note that translating this back to cofibration
sequences, this means that the connecting map \( \beta:\tilde{k}(2)_{10}X\r
BP\langle 2\rangle_{7}X
\)
satisfies
\(
\beta(\mu)=v_2a
\)
for some $a\in BP\langle 2\rangle_1X$ which, moreover, has non-trivial
reduction in $H_1(X,\Z)$. Note that this implies that \begin{equation}
\label{ahss6}v_1v_2(a)=0 \end{equation} (as $\beta$ is the connecting map of
$v_1$). Now we distinguish two cases. First, suppose $a$ is non-torsion.
Then there exists a cohomology class $u\in H^1(X,\Z)$ such that $\langle
a,u\rangle\neq 0$. But then (\ref{ahss6}) implies that $u$ cannot lift to
$BP\langle 2\rangle^1X$, which contradicts the fact that the $BP\langle
2\rangle$-cohomology AHSS collapses on $S^1$ (which represents $H^1(X,\Z)$).
The second case is that $a$ is torsion, but then there is a class $u\in
H^2(X,\Z)$ which is the Bockstein of a class $v\in H^1(X,\Z/m)$ such that
$\langle a\mod m, v\rangle\neq 0$. Similarly as above, however, we see from
(\ref{ahss6}) that $u$ cannot lift to $BP\langle 2\rangle^2X$, which is
again a contradiction, as $H^2(X,\Z)$ is represented by maps to 
$\C P^{\infty}$, and the $BP\langle 2\rangle$ cohomology AHSS collapses for
$\C P^{\infty}$. This concludes the proof of (\ref{ahss5}).

\vspace{3mm}
Therefore, we have shown that $X$ is orientable with respect to $BP\langle 2
\rangle$. But we have a comparison map of ring spectra \( BP\langle 2\rangle
\r E(2)
\)
(which is inclusion on coefficients), and therefore $X$ is
$E(2)$-orientable, as claimed.

\subsection{The $EO(2)$ story}
\label{elleo}

In this section, we will prove the following:

\vspace{3mm}
\noindent
{\em A Spin-manifold $X$ is orientable with respect to $EO(2)$ if and only
if it satisfies $w_4=0$.}

\vspace{3mm}
We first focus on the necessity of the condition. As remarked above, $EO(2)$
can be considered the spectrum of fixed points (or, in this case
equivalently, homotopy fixed
points) of the equivariant cohomology theory $E\R(2)$, which is a
$\Z/2$-equivariant version of $E(2)$. The $\Z/2$-equivariant real elliptic
cohomology spectrum was 
considered extensively in \cite{hk}. Now the primary obstruction 
to orientability with respect to $\Z/2$-equivariant integral 
Morava $K(2)$-theory $\tilde{K}\R(2)$ is 
\beg{q2m}{\mathbb{Q}_2[X]
}
where $\mathbb{Q}_2$ is the second real Milnor primitive
(more precisely, we should be considering again the integral version of the
Milnor primitive, but for the purposes of necessity, if the integral element
vanishes, so does its $\mod 2$ reduction). The story of the
$\Z/2$-equivariant Steenrod algebra is told extensively in \cite{hk},
Section 6. It turns out that the dual of this algebra can be described in
two equivalent ways: either
\beg{esteen1}{A_{\star}^{c}=\Z/2[\zeta_i,a,\sigma,\sigma^{-1}]}
where $dim(\zeta_i)=2^{i}-1$, $i=1,2,...$, $dim(a)=-\alpha$,
$dim(\sigma)=1-\alpha$ (see Appendix \ref{B}). The other description is as
\beg{esteen2}{A_{\star}^{c}=\Z/2[\xi_i,\tau_i,a,\sigma,\sigma^{-1}]}
where now $dim(\xi_i)=(1+\alpha)(2^i-1),\; i\geq 1$, 
$dim(\tau_i)=1+(1+\alpha)(2^i-1),\; i\geq 0$. In this second description
\rref{esteen2}, we then have a basis of the $\Z/2$-equivariant Steenrod
algebra $A^{\star}_{c}$ which 
is the $\Z/2[[a]][\sigma,\sigma^{-1}]$-basis dual to \beg{esteen3}{\prod
\tau_{i}^{\epsilon_i}\prod\xi_{i}^{r_{i}}}
(where $\epsilon_i=0,1$, $r_i=0,1,2,...$ and all but finitely many of these
values are $0$). This dual basis can be written as
\beg{esmilnor}{\prod\mathbb{Q}_{i}^{\epsilon_i}\prod
\mathbb{P}^{(r_1,r_2,...)}.} The elements $\mathbb{Q}_i$ (of dimension
$1+(1+\alpha)(2^i-1)$) are the real Milnor primitives. On the other hand,
using the description \rref{esteen1}, also the usual $p=2$-Milnor basis 
\beg{esnorm}{Sq^{R}}
serves as a
$\Z/2[[a]][\sigma,\sigma^{-1}]$-basis of $A^{\star}_{c}$.

\vspace{3mm}
Now the key to understanding the obstruction \rref{q2m} is converting the
element $\mathbb{Q}_2$ to the basis \rref{esnorm}. In particular, we claim
that one of the summands of $\mathbb{Q}_2$ is \beg{esq4}{Sq^{(4,0,...)}a^3.}
To show that is equivalent to expanding $\zeta_{1}^{4}$ in the basis
\rref{esteen3}. Using the recursive formulas given in \cite{hk}, Section 6, 
we find that 
$$\begin{array}{r}\zeta_{1}^{4}\equiv\tau_2
a^3+\xi_2\sigma^{-1}a^2+\xi_{1}^{2}\sigma^{-2}
+\xi_{2}\tau_{0}a^3 \\
+\xi_{1}^{2}\tau_{1}a^3 + \xi_{1}^{3}\sigma^{-1}a^2 +
\xi_{1}^{3}\tau_{0}a^{3} \mod a^4,\end{array}$$
so we see that $\tau_2 a^3$ is present, which proves that \rref{esq4} is
present in the expansion of $\mathbb{Q}_2$. This proves that
$a^3Sq^{(4,0,...)}[X]$ is a summand of \rref{q2m}, but for a Spin-manifold, 
$$Sq^{(4,0,...)}[X] = Sq^4[X],$$
which is Poincar\'{e} dual to $w_4$. This concludes the proof of necessity
of our condition.

\vspace{3mm}
Sufficiency follows from the advanced theory which Ando-Hopkins-Rezk 
\cite{ahr} developed to treat orientability with respect to $tmf$. We give
the argument here merely for completeness. One starts with the concepts of
May-Quinn-Ray \cite{mqr}: For an $E_{\infty}$ ring spectrum $E$ (a
multiplicative generalized cohomology theory with a particularly nice
multiplication), one can construct an infinite loop space of units of $E$
which \cite{ahr} 
denote by $GL_1(E)$ (\cite{mqr} use the notation $E^{\otimes}$). As a space,
(i.e. forgetting infinite loop space structure), $GL_1(E)$ is simply the
disjoint union of the connected components of the infinite loop space
associated with $E$ which are invertible under multiplication in $\pi_0(E)$.

\vspace{3mm}
Now infinite loop spaces correspond essentially bijectively with connective
spectra (infinite loop space theories whose coefficients are $0$ in negative
degrees). A convention of \cite{ahr} assigns to an infinite loop space
written in capital letters the connective spectrum with the same notation
but in lower case letters; for example, the connective spectrum
corresponding to $GL_1(E)$ is $gl_1(E)$. 

\vspace{3mm}
Now Rezk \cite{rezk}, following Kuhn \cite{kuhn}, defines a map
\beg{erezkln}{\phi: gl_1(E_2)\r E_2.}
(More generally, he defines for any $E_{\infty}$-ring spectrum $E$ a map
$gl_1(E)\r L_{K(n)}(E)$.) Now recall from the last paragraph that for $q>0$,

$$\pi_q(gl_1(E))\cong
\pi_q(E).$$
Therefore, one may ask in that notation what the map $\phi$ of
\rref{erezkln} induces on homotopy groups. Rezk gives a complete formula in
\cite{rezk}, which goes as follows (working at the prime $2$): For an
element 
$x\in (\widetilde{E_2})^0 S^{2k}$ (reduced $E_2$-cohomology), one has 
\beg{erezkln1}{
\phi_*(x)= x-\frac{1}{2}\cform{\sum}{i=1}{3}\psi_{\alpha_i}(x)+\psi_{2}(x)
}
where $\psi_{\alpha_i}$, $\psi_2$ are Ando's power operations \cite{ando}.
Before explaining them, we should make two remarks. First of all, Rezk
\cite{rezk} makes a general computation for any $(E_n)^0$-cohomology
element; we only give here the simplified formula needed for our purposes.
On the power operations, we also only use a simplified case, and simplify
notation for that purpose. In our notation, $\alpha_i$ are the three
solutions to the equation $$([2]_F\alpha)/\alpha=0$$ in some integral
extension of the ring $(E_2)^0$. 
For $z\in (E_2)^0(X)$ for a space $X$, the Ando power operation is then
$$\psi_{\alpha_i}(z)\in (E_2)^0[\alpha_i]$$ which is the image by $x\mapsto
\alpha_i$ of the element of $$(E_2)^{0}(X)[[x]]/[2]_F(x)= (E_2)^0 (X\times
B\Z/2)$$ which is given by the composition $$X\times B\Z/2 \r
E\Z/2\times_{\Z/2}X^2\r E\Z/2\rtimes_{\Z/2}(E_2)^{2}\r E_2.$$ Here the first
map is the diagonal, the middle map is induced by $z:X\r E_{2}$, the last
map is the power operation in $E_2$ given by the fact that $E_2$ is an
$E_{\infty}$-ring spectrum. Similarly, the operation $\psi_2(z)\in
(E_2)^0(X)$ is defined as the image of $x\mapsto \alpha_1$, $y\mapsto
\alpha_2$ of the element of $$(E_2)^{0}[[x,y]]/([2]_F(x),
[2]_F(y))=(E_2)^0(X\times B\Z/2\times B\Z/2)$$ which is given by the
composition $$\begin{array}{r}X\times B(\Z/2\times\Z/2)\r E(\Z/2\times\Z/2)
\times_{\Z/2\times\Z/2} X^4\\ \r
E(\Z/2\times\Z/2)\rtimes_{\Z/2\times\Z/2}(E_2)^{4}\r E_2 \end{array} $$
where the maps are analogous as above.

\vspace{3mm}
For our purposes, what matters is that for the periodicity element $u\in
(\widetilde{E_2})^0 S^{2}$, one has $$\psi_{\alpha_i}(u)=u(u+_F\alpha_i),$$
$$\psi_{2}(u)=u\cform{\prod}{i=1}{3}(u+_F\alpha_i)$$
while also $u^2=0$. Therefore, also using the fact that the Ando operations
give a homomorphism of rings, for the generator $u_k\in (\widetilde{E_2})^0
S^{2k}$, we get
$$\cform{\sum}{i=1}{3}\psi_{\alpha_i}(u)=u_k((\alpha_1)^k+(\alpha_2)^k+
(\alpha_3)^k),$$
$$\psi_2(u)=u_k(\alpha_1\alpha_2\alpha_3)^k.$$
These elements are in $(E_2)^0(S^{2k})$. To calculate them, note that 
$$[2]_F(x)=2x+_F a x^2 +_F x^4=(x-\alpha_1)(x-\alpha_2)(x-\alpha_3)(1+K_1
x+...)$$ (neglecting $u$ and recalling that $(E_2)^0=W_2[[a]]$ with an
element $a$ of dimension $0$ - this is not the element $a$ which occurs in
the $\Z/2$-equivariant Steenrod algebra), so
$$\alpha_1\alpha_2\alpha_3=-2,$$ $$\alpha_1\alpha_2
+\alpha_1\alpha_3+\alpha_2\alpha_3=a +2K_1.$$ We also know
$$\text{$\alpha_1+\alpha_2+\alpha_3$ is divisible by $2$},$$ because
otherwise \rref{erezkln1} would not be integral.
>From these equations, setting
$$s_k=(\alpha_{1})^k +(\alpha_2)^k +(\alpha_3)^k$$
and using Newton's formula, one gets
$$2|s_1=2a,\; s_2\in 2J,\; s_3\in 2J-6,$$
$$\text{$s_k\in 2J$ for $k>3$}$$
where $J$ is the ideal in $(E_2)^0$ generated by $2$ and $a$. Note that only
the summand $-6$ of $s_3$ is not in $2J$. From this, using \rref{erezkln1},
we get the following:

\vspace{3mm}
\noindent
{\em The only non-zero homotopy group of the fiber $F$ of the map $\phi$ of
\rref{erezkln} in dimension $\geq 3$ is in dimension $5$.}

\vspace{3mm}
Now $gl_1$ is a (topological) right adjoint functor from the homotopy
category of spectra to the homotopy category of connected spectra, so for a
finite group $G$ acting on a spectrum $E$, $gl_1(E^{hG})$ is the connective
cover of $(gl_1(E))^{hG}$ where $E^{hG}$ denotes the homotopy fixed point
spectrum of $E$ under the $G$-action. Therefore, the above assertion gives

\vspace{3mm}
\noindent
{\em For any action of the group $\Z/2$ on $E_2$,
the only non-zero homotopy groups of the fiber $F_{\Z/2}$ of the map
$\phi^{h\Z/2}:gl_1((E_2)^{h\Z/2})\r (E_{2})^{h\Z/2}$ in dimensions $\geq 3$
are in dimensions $\leq 5$. Moreover, the homotopy groups in dimensions $3$,
$4$ are $\Z/2$-modules.}

\vspace{3mm}
Now the theory of May, Quinn and Ray \cite{mqr} shows that 
for a map of spectra
$$sp\r gl_1(S)$$
(where $sp$ is some connective spectrum),
and an $E_{\infty}$-ring spectrum $E$, an $E_{\infty}$-orientation from the
corresponding Thom spectrum $Msp$ to $E$ $$Msp\r E$$ exists if and only if
the composition \beg{emqr}{sp\r gl_1(S)\r gl_1(E)} vanishes where the second
map is induced by the unit of $E$. 
For example, this can be applied with $sp=Spin$, $E=(E_2)^{h\Z/2}$ for the
finite group $\Z/2$ acting on $E_2$. But recall that the only non-zero
homotopy group of $Spin$ of dimension $<7$ is $\Z$ in dimension $3$;
therefore, by the above assertion, the map \rref{emqr} must factor through
the map $$\lambda:spin\r \Sigma^3 H\Z/2$$ which induces an onto map in
$\pi_3$. Therefore, the obstruction vanishes when pulled back to the fiber
$sp$ of $\lambda$. However, by general arguments of cobordism theory it
follows that the corresponding Thom spectrum $Msp$ classifies precisely
$Spin$-manifolds with $w_4=0$. Therefore, we have proved in particular that
all such manifolds are $(E_2)^{\Z/2}$-orientable.

\section{Physical interpretation: The connection between elliptic cohomology
and $M$-theory}
\label{eint}

In this section, we gather our evidence that elliptic cohomology is very
closely 
and fundamentally tied
to $M$-theory. This includes the evidence from the previous sections, as
well as other clues. We only outline the ideas as we plan to investigate the
proposals in this section in more detail and more carefully elsewhere
\cite{prog}. We continue to work with a $10$-dimensional compact
$Spin$-manifold $X$ and put $Y=X\times S^1$. Note that elliptic cohomology
of $X$ and $Y$ are related by the K\"{u}nneth theorem:
$E^*(Y)=E^*(X)\otimes_{E^*} E^{*}(S^1)$, so \( E^n(Y)=E^n(X)\oplus
E^{n-1}(X).
\)

\subsection{The twisting}
Now let us first discuss twisting. In the previous sections,
we focused on the untwisted case, but let us now briefly consider the case
when the background is twisted by an NS $3$-form $H$-field. In that case, as
noted in \cite{dmw}, Section 11, in the IIA part of the discussion, one must
replace $K$-theory by twisted $K$-theory, which was done in \cite{hm}.
However, Douglas \cite{dtwist} remarks that twistings of $K$-theory
determine topological modular forms: more precisely, twistings of $K$-theory
are classified by the space $BGL_1(K)$ and we have a map $BGL_1(K)\r tmf$.
Therefore, we see that twistings of $K$-theory are encoded in $tmf$, which
further maps into $E$. Hence, while in twisted $K$-theory we must alter the
theory with each $H$-field, elliptic cohomology unifies all of these twisted
cases in one theory. An explanation of this phenomenon may again lie in the
connection between elliptic cohomology and $2$-vector bundles, as discussed
in \cite{bdr}, \cite{hkdb}.

\vspace{3mm}
While $K$-theory twistings give rise to actual topological modular forms, we
have argued that complex-oriented elliptic cohomology seems to play a more
basic role in the present case. This hints that more ``twisting'' should be
allowed. 
Presumably, this 
should be the twisting by the field strength $H_7$, which is the field
strength associated with the $NS5$-brane, in a dual way to the situation
where $H_3$ is associated with the fundamental string $F1$. 
The lift to M-theory of the NS branes leads to the M-branes. The fundamental
string expands in one dimension along the M-theory circle to become the
M2-brane, whereas the $NS5$-brane lifts to M5 brane, thus maintaing the same
worldvolume dimension. The reason for 
this difference in codimension in the lifting is a consequence of the 
dimensions of the relevant forms. In 10 dimensions, the NS field strengths
have dimensions three and seven, while in eleven dimensions, the fields have
dimensions four and seven. 

\vspace{3mm}
The main result of \cite{dmw} shows, in the case $Y=X\times S^1$, a match
between the partition function of $M$-theory calculated from the
$G_4$-field, which is the field strength assoicated with the 
$M2$-brane, and the IIA-partition function, which is calculated from the
fundamental string RR sector. 
When
increasing the coupling in IIA, we get $M$-theory compactified on $X\times
S^1$, and in this duality, the fundamental string acquires another
dimension, and becomes identified with the $M2$-brane. This again suggests
that the elliptic refinement of the partition function which we proposed
reflects in some way interaction between the $M2$ and $M5$-brane. Note that
the $M5$-brane is an object in $M$-theory, with electro-magnetically dual
coupling to the field strength $G_4$. 
In IIA, this object loses one dimension, and becomes the $D4$-brane.
However, using the strong-weak duality, we may also, as above, identify 
$M5$ with
a $6$-dimensional object, namely the $NS5$-brane in IIA which couples
magnetically to the NS charge. From this point of view, the behaviors  
of $M5$ and $M2$ are symmetrical.

\vspace{3mm}
If we denote the worldvolume of the
$NS5$-brane by $W$, then the fundamental class $\kappa\in H_6(W)$ must
satisfy $Sq^3(\kappa)=0\in H_3(X,\Z)$, i.e. $\kappa$ must lift to the
$K$-theory 
homology of $X$. This is also $d^3$ in the $E(2)$-homology AHSS of $X$.
However, similarly as in Section \ref{elle} above, the next differential
$d^5$ lands in homological dimension $1$ and hence is excluded, so $\kappa$
lifts to a class in 
$E(2)_6(X)$. Now by multiplying by the elliptic cohomology periodicity
element $v_{2}^{-1}$, which is in dimension $-6$, we get an element of
$E(2)_0(X)$.

\vspace{3mm}
It is worth commenting that $6$ is the only dimension of a world volume
$\leq 10$ which can be shifted to $0$ by inverting the element $v_2$. Note
that, unlike in $K$-theory, the Bott element $v_1$ is not inverted in
elliptic cohomology: this may be singling out the $5$-brane as the object
whose interactions 
give the main part of the $M$-theoretical correction to the IIA partition
function. In $M$-theory, the $M5$-brane couples magnetically to the field
strength $G_4$. In \cite{dmw}, Section 7.2, arguments are given pointing out
the naturality of a choice of coordinates under which $G_4$ gives the main
contribution to their partition function (the approximation with which we
are working here).
>From another point of view, in type II string theory,
RR $D$-branes of lower dimension can be generated from higher dimension by
tachyonic condensation. This involves the Gysin isomorphism, i.e. 
the Bott element. In $M$ theory, as not all even dimensions of branes are
allowed (specifically, we cannot reverse the process and for example turn a
$2$-brane into a $4$-brane in $M$-theory). This, again, seems related to the
non-invertibility of the element $v_1$ in elliptic cohomology.

\vspace{3mm}
But what might be the twisting with respect to the field strength $H_7$? For
this, note that $v_2\in MU_{*}$ is represented by a complex manifold, namely
the {\em Milnor manifold} whose Segre characteristic number is $2$. However,
an $M5$-brane, while it must be orientable, may not be a complex manifold.
Therefore, it is reasonable to propose that inclusion of such non-complex
$M5$-branes will introduce a new twisting.

\vspace{3mm}
Elliptic cohomology has the formal group law of an elliptic curve. Hence one
should be able to see the group (or possibly subgroups of) $SL(2,\ZZ)$.
Since the modular 
parameters that appear in M-theory and string theory are usually of the form
(e.g. \cite{sen, ganor})
\(
\tau= ({\rm field}) + i({\rm volume~modulus})
\)
it seems reasonable to propose the moduli in the form  
\begin{eqnarray}
\tau_2= \langle B_2, [\Sigma_2] \rangle + i vol(\Sigma_2) 
\\
\tau_6= \langle B_6, [\Sigma_6] \rangle + i vol(\Sigma_6) \end{eqnarray}
where $\langle ~,~ \rangle$ is the Kronecker product, $[~~]$ is the
fundamental class, $\Sigma_2$ and $\Sigma_6$ are two- and six-cycles that
can correspond to $F1$ and $NS5$ respectively. The modular parameter
$\tau_6$ should be related to the map \rref{char} by an equation which we
can schematically write as $$q=e^{2\pi i \tau_6}.$$ Note however that in
order to make physical predictions from this, we would need a more precise
normalization of coordinates to predict the exact choice of group of modular
transformations (see also the comments on elliptic spectra in Appendix
\ref{B}), as well as a formula for the $M5$-brane charge (see next Section).

\subsection{$M2$ and $M5$-branes}

It remains to give an interpretation of $v_2^{-1}v_1^3$.
One possible suggestion in the corresponding M-theory picture is that the
$M2$ and the $M5$ coexist and that 
the membrane modulus is not inverted. The first point can be understood from

M-theory as the general statement that if a soliton spectrum contains the 
$M5$-brane then it automatically contains the $M2$-brane. This can be
understood by the Hanany-Witten effect that implies that an $M2$-brane is 
created when two fivebranes cross. Alternatively, an $M2$-brane appears from
dielectric $M5$-branes 
in the limit when the 3-cycle shrinks to zero. 

\vspace{3mm}
Let us review briefly the intersections and bound states of M-branes. M2 and
M5 can consistently coexist, 
in compatability with the gravitational anomaly cancellation. They obey a
Dirac quantization condition 
$eg=2\pi n G_N$ where $e$ is the M2 charge and $g$ is the M5 charge and
$G_N$ is the eleven-dimensional Newton 
constant. The presence
of M2 and M5 modify the equation of motion to
\footnote{Such a modification has been discussed in \cite{Lech} using
Chern-kernels.} \footnote{ A more careful 
analysis would perhaps involve more refined treatment like 
 Cheeger-Simons differential characters \cite{dfm}. As we mentioned earlier,
we will 
investigate the ideas of this section more carefully in \cite{prog}.} 
\( d*G_4=G_4 \wedge G_4 + g f_3 \wedge J_5 + e J_8 + \frac{2 \pi G_N}{g}X_8
\)
where $J_5$ is the current (poincar\'e dual) of M5, and $J_8$ that of M2.
$X_8$ is the 8-polynomial  associated with the gravitaional anomaly, and
$f_3$ is the $M5$ worldvolume gauge field.
>From supergravity, the allowed supersymmetric intersections (see e.g. 
\cite{jerome}
for a review) include:
\footnote{notation: $(p)$ means the intersection is over a p-brane. Note
that these are orthogonal intersections.} $M2 \cap M2 (0)$, $M2 \cap M5
(1)$, $M5 \cap M5 (1)$, and 
$M5 \cap M5 (3)$.
The second is especially interesting because it is the intersection over a
string and 
is one of the building blocks for
brane-intersections. The reduction of the intersection $M2 \cap M5 (1)$
along M2 leads to F1 ending on NS5, and leads to D2 ending on D4 if the
reduction is along one of the coordinates of M5. There are many possible
D-brane bound states in type II string theory. One way they arise is by
placing D-branes in constant background B-field. The worldvolume coordinates
of the Dp-brane become noncommutative (=NC) along the directions of the
non-vanishing B-field. If B is spacelike, one can define a decoupling limit
of NCYM, i.e. a NC field theory. If B is timelike, one gets noncommutative
open string theory (NCOS). In principle, any bound state in type IIA should
have a lift to M-theory, and the analogue situation is M-branes in
(constant) background $C$-field. The configuration of M2 branes ending on M5
is the lift 
of strings ending on D-branes.
One can similarly define a decoupled theory, the light Open Membrane (OM)
theory \cite{om}. Six-dimensional OM-theory is the high energy limit of
5-dimensional NCYM and NCOS. Compactification of OM-theory on an electric
(resp. magnetic) circle leads to NCOS (resp. NCYM). A constant background
C-field can be traded for a constant M5 worldvolume field $f_3$. This
represents a bound state of this M5 with a delocalized 
M2 along 2 of the 5 spatial directions of M5.
Many of the bound states involving different combinations can be related to
$(M2, M5)$ by a (Lorentz) transformation, and this seems to indicate that
the latter is the ``basic'' bound state.

\vspace{3mm}
Perhaps even more importantly
for us, however, this suggests that we  
may use the IIA-side of the \cite{dmw} calculation to interpret the elliptic
refinement, as the $G_4$ approach should be exactly dual. The beginning of
such argument was seen at the end of the last section. But the IIA-side is
easier to work with, as we have at our disposal the usual expansion of the
radial excitation modes of a fundamental string. As we already saw, the
suggestion about the meaning of the parameter $v_{2}^{-1}v_1^{3}$ is that it
comes from interaction with a complex oriented $M5$-brane, and that the
partition function should be twisted if the $M5$-brane is not
complex-oriented. Therefore, we should consider what kind of possible bound
states between $M2$ and $M5$-branes can arise. One striking suggestion is an
open $M2$-brane with boundary on the $M5$-brane. One feature which could
suggest those states is the non-invertibility of $v_1$, and hence the
element $v_{2}^{-1}v_1^{3}$, in elliptic cohomology. However, from the point
of view of the $M5$-brane, it is not clear whether such states would not be
anomalous. Also, in the IIA dimensional reduction, we are not seeing any
direct role of the open string partition function. It may therefore be that
rather than ending on the $M5$-brane, the $M2$-brane {\em intersects} the
$M5$-brane in a fundamental string, and the elliptic partition function
reflects the energy such bound state acquires from the intersection. One
might also argue that in analogy to the string theory situation where the 
open sector requires the existence of the closed sector, the open membrane 
requires the existence of the closed membrane. However, the open states 
seem to be only needed in this setting to imply the existence of the branes,
and they do not enter the calculations of the partition function.
\footnote{In \cite{dmw,ms,hm}, the branes are not directly used in
calculating the partition function, and in discussing them, we are assuming
further that one can include them and use them for the spacetime result.} 

\vspace{3mm}
We therefore conjecture that the elliptic refinement of the IIA- (and
alternately the M-theory $G_4$-) partition function picks up states arising
from intersection of an $M2$-brane with an $M5$-brane, which could be
$H_7$-twisted if the $M5$-brane is not complex-oriented. In the present
(untwisted) case, there is an anomaly of these states when $w_4\neq 0$. On
the IIA side for $Y=X\times S^1$ (the case considered here), the
intersection is the (end of the) IIA fundamental string. The elliptic
partition function which we constructed, after suitable normalization and
computation of the $5$-brane charge should compute the 
the $M2$-$M5$-intersecting state correction to the $G_4$ M-theory partition
function.

\vspace{3mm}
In section \ref{elleo} we proved that orientability with respect to $EO(2)$
is equivalent 
to vanishing of $w_4$. We would like to point out the relevance of this to
the M-theory backgrounds as well as to the $M5$ anomaly. In \cite{flux},
Witten showed that $G_4/{2 \pi}$ is quantized as $w_4/2$ mod $\ZZ$. The the
condition $w_4=0$ implies that there are no half-integral fluxes, which is
the case for the relevant $\ZZ_2$-orbifolds (and orientifolds). Later in
\cite{duality}, Witten showed that $w_4$ also shows up as (part of the) mod
2 index and, 
consequently, the anomaly for the $M5$-brane. We will revisit this in
\cite{prog} 
and study possible relation between $1/2$-integral fluxes and twisting by
$H_7$ in this context.

\subsection{Closing remarks}
There is also a purely mathematical side of
these phenomena:
In \cite{hkconf}, a model of elliptic cohomology was proposed based on
$\mathcal{E}$-equivariant {\em stringy bundles} over an elliptic curve
$\mathcal{E}$. A stringy bundle is a variant of what \cite{segal, st} call
an {\em elliptic object}, i.e. a ``conformal field theory indexed by
$\mathcal{E}$''. 
The authors of \cite{hkconf} could not figure out what was the role of the
elliptic curve $\mathcal{E}$ in the ``spacetime'' part of the theory. But
the present context suggests that this may be perhaps interpreted as an
intersection between an $M2$-brane and an $M5$-brane, and that the roles of
these $M$-branes should be further investigated to enhance geometric
interpretation of elliptic cohomology. It should be emphasized that in 
\cite{hkconf}, there was no object which would play the
role of the spacetime manifold $X$, as the mathematical story is concerned 
with abstract CFT only, and not superstring theory of a given type. However,
the present observations may suggest that even there, interactions of the
$M2$-brane and $M5$-brane may play a role.

The physical interpretation in this section deserves to be studied more
carefully and 
in more detail, in particular on the topics of normalization and
identification of modular group, as well as an $M5$-brane charge formula. We
hope to achieve this in \cite{prog}.

M-theory continues to prove how rich it is both physically and
mathematically. 
We hope that elliptic cohomology and Morava K-theory could be ``derived'' 
in the future from M-theory, at the level of partition functions, in the
sense of \cite{dmw} for K-theory and help in completing the derivation 
initiated in \cite{hm} for twisted K-theory. 

\bigskip\bigskip
\noindent

\section*{Acknowledgements}
\noindent
H.~S. would like to thank V.~Mathai for useful discussions on \cite{dmw}. He
also thanks the Michigan Center for Theoretical Physics for hospitality, 
during which this project was started. The authors would like to thank
Matthew Ando, Mike Hopkins and Charles Rezk for discussions, especially on
their theory of $tmf$-orientability. 

\noindent
The first author's research was supported by NSF grant DMS 0305853.
The second author's research was supported by the Australian Research 
Council.

\newpage
\setcounter{section}{0}
\renewcommand{\thesection}{Appendix \Alph{section}}
\setcounter{equation}{0}

\begin{appendix}

\section{Appendix: Orientability} 
\label{A}

In this appendix, we collect some definitions and basic properties of
orientation \cite{rud}. Let us start with the simplest case. The orientation
of $\RR^n$ can be defined homologically as one of the generators of the 
group $\ZZ=H_n(\hat{\RR}^n,\ZZ)=H^n(\hat{\RR}^n,\ZZ)$ where
$\hat{\RR}^n=S^n$ is the one-point compactification.

If X is a closed connected manifold with $H_n(X,\ZZ)=\ZZ$ then every
generator $[X]$ 
of $H_n(X,\ZZ)$ can be considered as an orientation of $X$.

A $\RR^n$-bundle $\xi$ over a connected manifold is orientable if 
$H^n(T\xi,\ZZ)=\ZZ$, where $T\xi$ is the Thom space of $\xi$. An orientation
of 
$\xi$ is a generator of $H^n(T\xi,\ZZ)$. 

For an arbitrary cohomology theory $E$, an $E$-orientation of a closed
manifold $X^n$ is an element $[X]\in E_n(X^n)$.

Examples of orientation:
\begin{enumerate}
\item {\bf $H\ZZ$-orientability:} A vector bundle $\xi$ is $H\ZZ$-orientable
if and only if 
its structure group can be reduced to $SO$. This holds if and only if
$w_1(\xi)=0$. By using the Thom isomorphism $w_1(\xi)=\phi^{-1}Sq^1u_{\xi}$,
this is equivalent to $Sq^1(u)=0$, where $u=u_{\xi}\in H^n(T\xi,\ZZ_2)$.
\item{\bf KO-orientability:} A vector bundle $\xi$ is $KO$-orientable if and
only if it admits a spin structure, i.e. if and only if
$w_1(\xi)=0=w_2(\xi)$. \item{\bf K-orientability:} A vector bundle  $\xi$ is
$K$-orientable if and only if it admits a ${\rm spin}^c$ structure, i.e. if
and only if $w_1(\xi)=0=\beta w_2(\xi)$. Note that every $KO$-orientable
vector bundle is also $K$-orientable. \item{\bf E-orientability:} Let $X$ be
a topological manifold. An element 
$[X,\del X]\in E_n(X,\del X)$ is an $E$-orientation if 
\(
{\cal{E}}_*^{m,U}[X,\del X]=\pm s_n
\)
for every $m\in X$ and every disk neighborhood of $m$. Here 
${\cal{E}}^{m,U}: X \r S^n$ is the map that collapses the complement of $U$.
$s_n \in E_n(S^n,*)$ is the image of $1\in \pi_0(E)$ under the isomorphism
\(
\pi_0(E)=\tilde{E}_0(S^0)\cong\tilde{E}_n(S^n)=E_n(S^n,*)
\)
and is a canonical orientation of the sphere $S^n$.
A closed manifold is $E$-orientable iff its stable normal bundle is 
$E$-orientable.
\end{enumerate}

Oriented objects have a lot of good properties: \begin{enumerate} \item
There is a Thom-Dold isomorphism $\phi: E^i(X) \r \tilde E^{i+n}(T\xi)$ for
every $E$-oriented $\RR^n$-bundle $\xi$ over X. \item There is Poincar\'e
duality: $E^i(X) \r E_{n-i}(X,\del X)$ for every $E$-oriented manifold
$X^n$. \item One can generalize the classical Chen classes and develop a
theory of characteristic classes taking values in $E^*$ provided all complex
vector bundles are $E$-orientable. Those are called Chern-Dold classes.
\item They provide integral invariants. Start with any $D$-orientable
manifold $X^n$, having $D$- and $E$- orientations $[X]_D$ and $[X]_E$,
respectively.  
Let $\tau : D \r E$ be a ring morphism of ring
spectra. Let $\xi$ be any $D$-orientable (and hence $E$-orientable) object
over $X$, and let $u_D$ (rep. $u_E$) be a $D$- (reps. $E$-) orientation of
$\xi$. Integrality 
phenomena arise because of incompatability of the orientations, i.e. $\tau
(u_D)\neq u_E$. The orientation $u_E$ gives rise to the Thom-Dold
isomorphism 
$\phi_E: E^*(X)\r {\tilde E}^*(T\xi)$. Set 
\(
R(\xi)=R_{u_D,u_E}(\xi):=\phi_E^{-1}\tau (u_D) \in E^0(X)
\)
The orientations $[X]_D$ and $[X]_E$ determine orientations 
$u_D(\nu)$ and $u_E(\nu)$ of the stable normal bundle of $X$. Then, one can
write equate Kronecker pairings using the two different orientations \(
\langle \tau(x) R(\nu), [X]_E \rangle = \tau \langle x, [X]_D \rangle
\label{integral}
\)
The integrality theorem says that the element $\langle \tau(x) R(\nu), [X]_E
\rangle$ 
of the group $\pi_{n-k}(E)$ belongs to the subgroup 
$Im\{ \tau_*: \pi_{n-k}(D) \r  \pi_{n-k}(E) \}$.

\item Gysin homomorphism: Let $F:X^m\r Y^n$ be a map of closed manifolds. 
If both $X,Y$ are E-oriented, then one can define the Gysin homomorphisms:
\( f^{!}:F^i(X)\r F^{n-m+i}(Y), \quad f_{!}:F_i(X)\r F_{n-m+i}(Y)
\)
to be the compositions
\begin{eqnarray}
f^{!}:F^i(X)&\cong& F_{m-i}(X) {\buildrel {f_*} \over \longrightarrow}
F_{m-i}(Y)\cong F^{n-m+i}(N), ~~ f^{!}=P_{[Y]}^{-1}f_{*}P_{[X]} \nonumber\\
f_{!}:F_i(Y)&\cong& F^{n-i}(X) {\buildrel {f_*} \over \longrightarrow}
F^{n-i}(X)\cong F_{m-n+i}(Y), \quad f_{!}=P_{[X]}f^{*}P_{[Y]}^{-1} \nonumber
\end{eqnarray} If $Y$ and/or $X$ are not $E$-orientable but the difference
of normal bundles 
$\nu_Y-f^*\nu_X$ is, then one can still define similar maps.
 
In particular, if $m=n$, the manifolds are $H\ZZ$-oriented and the map has
degree $\pm1$ 
then $Y$ is $E$-orientable if $X$ is.
\end{enumerate}

The obstruction to orientability is given by some invariants called the
Postnikov invariants. For our purposes, an object is $E$-orientable iff $0
\in \kappa_n$ for all $n$, where 
$\kappa_n \in H^{n+1}\left(E_{(n+1)}; \pi_n(E)\right)$ is the $n$-th
Postnikov invariant of 
$E$.


\section{Appendix: Formal group laws, 
Milnor primitives and a menagerie of generalized
cohomology theories} 
\label{B}
\setcounter{equation}{0}

We recall here very briefly the story of formal group laws (FGL's), and how
they are relevant to 
complex-oriented generalized cohomology. The reader may refer to Appendix 3
of \cite{rav} for more 
detailed information. What we mean by a ($1$-dimensional, commutative) {\em
formal group law} over a 
(super)-commutative ring $R$ is a series \begin{eqnarray} &F(x,y)=x+_F y=
\sum a_{ij}x^i y^j\in R[[x,y]]&
\end{eqnarray}
which satisfies
\begin{eqnarray}
x+_F 0&=&x,
\nonumber \\
x+_F y&=&y+_F x,
\nonumber \\
(x+_F y)+_F z&=& x+_F (y+_F z).  
\end{eqnarray}
The significance in homotopy theory is as follows: we call a ring-valued
generalized 
cohomology theory $E$ {\em complex-oriented} if the canonical complex line
bundle $\gamma$ over $\C P^{\infty}$ is $E$-oriented, i.e. satisfies an
$E$-Thom isomorphism. A choice of such isomorphism is called a {\em complex
orientation} of $E$. It then follows that every complex bundle is
$E$-oriented. Moreover, the complex orientation determines a {\em Chern
classes} of complex bundles 
precisely analogously as in the case of ordinary cohomology. The way FGL's
enter the picture is that there exists a 
{\em unique} FGL over the ring $E_*$ (depending only on $E$ and complex
orientation) such that for complex line 
bundles $L$, $M$ over a space $X$, \begin{eqnarray}&c_1(L\otimes
M)=c_1(L)+_Fc_1(M).&
\end{eqnarray}
One should note that if we change the complex orientation on $E$, the formal
group law gets replaced by a 
{\em strongly isomorphic} FGL $G$, which means that there exists a formal
series 
$f(x)=x+a_2 x^2+ a_3 x^3+...\in E_*[[x]]$ such that 
\begin{eqnarray}
&f(x)+_G f(y) =f(x+_F y).&
\end{eqnarray}
The formal group law up to strong isomorphism is a powerful invariant of 
a complex-oriented generalized cohomology theory. The FGL of ordinary
cohomology is additive, i.e. $x+_F y=x+y$, and the FGL of $K$-theory is
multiplicative, i.e. $x+_F y =x+y+v_{1}xy$ where $v_1$ is the Bott element.
It should be noted that over a $\Q$-algebra, all FGL's are isomorphic, just
as all rational generalized cohomology theories are equivalent: this is a
manifestation of the fact that homotopy theory is, very much, ``the science
of torsion''. There exists a {\em universal} formal group law, i.e. a
commutative ring $L$ with a formal group law $F$ such that for any
(super)-commutative ring $R$ and any FGL $G$ on $R$ there is a unique
homomorphism of rings $L\r R$ which carries $F$ to $G$. This ring $L$ was
discovered by Lazard. Quillen discovered that, miraculously, $L$ is
isomorphic to the ring of coefficients of $MU$, complex cobordism, which is
in some sense the universal complex-oriented generalized cohomology theory!

\vspace{3mm}
The story goes onwards from there. If we localize at a prime $p$ (in this
paper, the relevant case is $p=2$), then $MU$ breaks up as a direct sum of
certain theories called $BP$ after Brown-Peterson \cite{bp}. There is a
corresponding notion of $p$-typical FGL, which however we shall not need to
consider. In any case, the coefficients of $BP$ are \(
BP_*=\Z[v_1,v_2,v_3,...],\;\;dim(v_n)=2p^{n}-2.
\)
The formal group law on $BP_*$ (the universal $p$-typical FGL) was also
first observed in algebra and is due to Cartier. By work of Baas-Sullivan
(with substantial recent 
improvements and simplifications), it is possible to create generalized 
cohomology theories at will by killing regular sequences in the ring $BP_*$
and/or inverting elements in its coefficient ring. This leads to a
``menagerie of generalized cohomology theories''. Some of the important
theories are the Johnson-Wilson theory $BP\langle n\rangle$ with
coefficients $\Z[v_1,...,v_n]$ and corresponding theory with $v_n$ inverted
which is the Landweber theory denoted by $E(n)$ (hence,
$E(n)_*=\Z[v_1,...,v_n,v_{n}^{-1}]$). One may also kill all of the lower
$v_n$'s to get the {\em integral Morava $K$-theory} $\tilde{K}(n)$ with
coefficients $\Z[v_n,v_{n}^{-1}]$, or kill also $p$ to get the {\em Morava
$K$-theory} $K(n)$ with coefficients $\Z/p[v_n,v_{n}^{-1}]$. There are also
{\em connective} versions of the Morava theories $\tilde{k}(n)$, $k(n)$ with
coefficients $\Z[v_n]$, $\Z/p[v_n]$, respectively.

\vspace{3mm}
Some of these theories have notable formal group laws. For example, the
formal group law on $K(n)_*$ has {\em height $n$}, which means that \( [p]_F
x =\underbrace{x+_F...+_F x}_{\text{$p$ times}}=v_n x^{p^n}.
\)
Studying isomorphisms of this FGL led Lubin and Tate \cite{lt} to the
discovery of the Lubin-Tate FGL on the ring
$W_n[[a_1,...,a_{n-1}]][u,u^{-1}]$ where $W_n$ is the ring of Witt vectors,
i.e. integers of an Eisenstein extension of order $p^n$ of the field of
$p$-adic numbers $\Q_p$. The Lubin-Tate coefficient ring turns out to be a
completed sum of copies of $E(n)_*$, and one can consider a generalized
cohomology theory with the Lubin-Tate coefficient ring, which is essentially
a completed sum of copies of $E(n)$. However, the reason Lubin-Tate laws
were invented has nothing to do with homotopy theory; rather, they are
needed for local class field theory (algebraic number theory).

\vspace{3mm}
For $n=2$, one can also consider cohomology theories whose formal group laws
are elliptic, i.e. are obtained by Taylor expansion of the group law on an
elliptic curve over some commutative ring. Such theories are called {\em
complex-oriented elliptic cohomology theories}. To be more precise about
this, Ando, Hopkins and Strickland \cite{tmf} define a {\em $2$-periodic}
ring spectrum as a generalized cohomology theory $E$ with an orientation of
the identical complex line bundle on $\C P^{\infty}$ which, when restricted
to $\C P^1$, has an inverse in $E_*$. If in addition $E_{2n+1}=0$ for all
$n$ (an {\em evenness condition}) and there is an elliptic curve
$\mathcal{E}$ over $E_*$ with an isomorphism of its formal group law with
the FGL of $E$, \cite{tmf} call $E$ an {\em elliptic spectrum}. (Note: the
word ``spectrum'' is a term of algebraic topology which means a generalized
cohomology theory. The reason for using a separate word is that spectra may
be refined to a 
category which contains {\em rigid} point set level information, analogous
to point set level maps of spaces. Cohomology theories, on the other hand,
see only maps ``up to homotopy''. The rigid level is 
necessary for certain more complicated constructions, such as (co)simplicial
realizations.)

\vspace{3mm}
Among the examples given in \cite{tmf} is the elliptic spectrum
$$E_*=\Z[a_1,a_2,a_3,a_4,a_6][u,u^{-1}]$$
associated with the Weierstrass curve which we write in the form
\beg{eweier}{y^2+a_1 uxy +a_3 u^3 y = x^3 +a_2 u^2 x^2 +a_4 u^4 x +a_6 u^6.}
Here we put $dim(x)=4$, $dim(y)=6$. The map (\ref{char}) for this spectrum
can be constructed using the so called Tate parametrization. Letting
$\sigma_k(n)=\cform{\sum}{d|n}{}d^k$, $\alpha_k=\cform{
\sum}{n>0}{}\sigma_k(n)q^n$, one can define (\ref{char}) by \begin{equation}
\label{etate} a_1=1, \;a_3=0,\;a_2=0,\; a_4=-5\alpha_3, \;
a_6=-(5\alpha_3+7\alpha_5)/12. \end{equation} (The coefficients of $a_6$
turn out to be integers.)

\vspace{3mm}
On the other hand, the Lubin-Tate spectrum
\beg{eslt}{E_*=W_2[[a]][u,u^{-1}]}
is also an elliptic spectrum, its corresponding elliptic curve being
\beg{eelt}{y^2+auxy+u^3y=x^3} (see \cite{hopm}). To construct the map
(\ref{char}) in this case (where the target is $K$-theory with coefficients
in $W_2$), the Lubin-Tate FGL has \beg{eltt}{[2]_Fx=(2x)+_F (aux^2) +_F (u^3
x^4)} and Lubin-Tate theory implies that over a ring where $au$ is
invertible, this is isomorphic to the multiplicative FGL. Thus, if we define
(\ref{char}) by $$au=v_1,\; u=v_1 q^{-1}$$ or equivalently $$a=q,\; u=v_1
q^{-1},$$ we get the correct map after composition with the automorphism of 
$(K\otimes W_2)[[q]][q^{-1}]$ which sends the FGL (\ref{eltt}) to the
multiplicative 
FGL of $K$-theory. We see that (\ref{eelt}) is
a Weierstrass curve with 
$$a_1=a,\; a_3=1,\; a_2=a_4=a_6=0,$$
but the map (\ref{char}) we constructed for this theory is however not the
same as the map coming from the Tate parametrization, thus further
confirming our observation about non-uniqueness of coordinates. Another
disadvantage of the theory \rref{eslt} is that the character map requires us
to use $K\otimes W_2$ and hence the Chern character involves crystalline
cohomology.

\vspace{3mm}
Regarding maps between the generalized cohomology theories mentioned, 
a pretty good guideline is usually what maps there are on coefficient rings
(without renaming elements). Thus, there are maps from $BP$ to all the
theories mentioned, and reductions from $BP\langle n\rangle$ to
$\tilde{k}_i$ to $k_i$, $1\leq i\leq n$. There are also maps $BP\langle
n\rangle\r E(n)$ and $\tilde{k}_n\r \tilde K(n)$, $k(n)\r K(n)$. There are
{\em no} maps between the $K(n)$'s for different $n$. 
However, there is an elliptic curve called the Tate curve, which can be,
formally at least, described as ``$\mathbb{G}_{n}/(q^{\Z})$'' where
$\mathbb{G}_m$ is the multiplicative group. This Tate curve 
has coefficients (domain of definition)
$\Z[[q]][q^{-1}]$ and its formal group law therefore is multiplicative. We
therefore have character maps such as (\ref{char}) from elliptic cohomology
theories to $K[[q]][q^{-1}]$ or its suitable completion.

\vspace{3mm}
An important part of stable homotopy story is {\em locality} with respect to
generalized cohomology theories, in the sense of Bousfield. We say that a
spectrum $X$ is {\em acyclic} with respect to a generalized cohomology
theory $E$ if $E^*X=0$. We say that a spectrum $Y$ is {\em $E$-local} if
$Y^*X=0$ whenever $X$ is $E$-acyclic. Bousfield \cite{bous} constructed an
$E$-localization map 
$$X\r L_E(X)$$
which is a universal map (in the homotopy sense) from $X$ to an $E$-local
spectrum. 
One should mention that localization with respect to the Moore spectrum 
$M\Z/p$ (the cofiber of the degree $p$ map $S^0\r S^0$) is the right notion
of $p$-completion of spectra. For many considerations in homotopy theory,
such completion at a prime $p$ is understood throughout without being
explicitly mentioned: this is what homotopy theorists mean by {\em working
at a prime $p$}. Now one distinction between the elliptic cohomology
theories mentioned above is behavior with respect to localization. For
example, the spectrum $E_2$ is $K(2)$-local (recall that $K(2)$ is the
second Morava $K$-theory). On the other hand, the spectrum $K[[q]][q^{-1}]$
is $K$-local). The spectrum $tmf$ is $E(2)$-local.

\vspace{3mm}
In homotopy theory, the story behind complex-oriented cohomology theories is
largely governed by the {\em Milnor primitives}. Milnor \cite{mil} found
that the dual of the Steenrod algebra $A_*$ can be simply described as
follows (we shall work at $p=2$): there is a comultiplication
$\psi:H^*(X,\Z/2)\r H^*(X,\Z/2)\hat{\otimes} A_*$ 
for any space
$X$ ($\hat{\otimes}$ denotes completed tensor product). Now one has for the
generator $a\in H^1(\R P^{\infty},\Z/2)$, \( \psi(a)=\sum a^{2^{i}}\otimes
\zeta_{i}
\)
and, in fact, $A_*=\Z/2[\zeta_1,\zeta_2,...]$, 
$dim(\zeta_i)=2^i-1$. So, one can obtain a basis of
$A^*$ by dualizing monomials in the $\zeta_i$'s. It turns out, however, that
the beneficial way of doing this is to dualize the basis 
\begin{equation}
\label{milnor1}
\zeta_{1}^{\epsilon_1}...
\zeta_{n}^{\epsilon_n}...\zeta_{1}^{2r_1}...\zeta_{n}^{2r_n}...
\end{equation}
Here $\epsilon_i\in\{0,1\}$ $r_i\in\mathbb{N}$ and obviously only finitely
many of the $\epsilon_i$'s and $r_i$'s are allowed to be non-zero. The
reason to write the basis in the awkward way (\ref{milnor1}) is that the
dual basis is then of the form \(
Q_{0}^{\epsilon_1}...Q_{n-1}^{\epsilon_n}...P^{(r_1,r_2,...)}
\)
where the $Q_i$'s form an exterior subalgebra of $A^*$. These elements are
called the Milnor primitives. The Milnor basis is different from the
Cartan-Serre basis, but the Adem relations lead to a recursive conversion
formula between both bases. It is not difficult to work out the conversions
explicitly in low dimensions, where there are not many elements to consider.

\vspace{3mm}
Now the connection with complex-oriented generalized cohomology theories is
as follows: Brown-Peterson \cite{bp} found that the primary Postnikov
invariants (the ones not attached to lower ones) of $BP$ are precisely the
Milnor primitives $Q_n$, and moreover the invariant $Q_n$ attaches precisely
the homotopy class $v_n$. 
Since we saw that all of the other theories we wrote
down receive maps from $BP$ which preserve names of generators, similar
conclusions hold for all of the theories involved, in particular Morava
$K$-theories. We therefore conclude that $Q_n$ is for example the primary
differential of the $K(n)$-AHSS (in homology or cohomology), and its
integral lift is the primary differential in the $\tilde{K}(n)$-AHSS, etc.
These facts we are using in the present paper.

\vspace{3mm}
There are refinements beyond the complex-oriented story which one needs to
consider. In particular, one can study {\em real-oriented generalized
cohomology theories} (see \cite{hk}), generalizing orthogonal $KO$-theory,
or more precisely, Atiyah's real $KR$-theory. 
The story is substantially more complicated there,
but the basic guideline is that there are real-oriented analogues to all the
complex oriented theories considered above. (Caution: the real-oriented
analogue of $MU$ is {\em not} $MO$ but Landweber's real cobordism $M\R$.)
The right way of considering real-oriented generalized cohomology theories
is as $\Z/2$-equivariant theories. This means that homology and cohomology
gets doubly indexed, i.e. by $k+\ell\alpha$ where $k,\ell\in\Z$, 
and $\alpha$ is the sign representation of $\Z/2$. There is always a
forgetful map from a real to the corresponding complex theory, which just
sends $\alpha$ to $1$. Now the main point is that complex bundles $\eta$
with a real structure (i.e. isomoprhism $\overline{\eta}\cong\eta$) are
orientable with respect to real-oriented generalized cohomology theories,
and the orientation lies in dimension $n(1+\alpha)$ where $n$ is the
dimension of the bundle. The formal group law story is thus repeated from
the complex case, but the calculation of coefficients is substantially more
difficult (many cases were worked out in \cite{hk} and in subsequent
papers).

\vspace{3mm}
One case which is particularly relevant here is the theory $EO(2)$ which is
the fixed point part of the theory $E\mathbb{R}(2)$, the real version of
$E(2)$. (Caution: Hopkins-Mahowald \cite{hopm} use symbols such as $EO(2)$,
and in particular $EO_2$ in a different meaning than \cite{hk}, namely to
denote fixed point spectra of $E_2$ under a larger finite group.) The
coefficient ring $E\mathbb{R}(2)_{\star}$ (here the symbol $\star$ means
that we are considering the general dimension $k+\ell\alpha$, see \cite{hk})
is a $\Z[a]/(2a)$-module where $a$ has dimension $-\alpha$. The generators
of this module are monomials (let us call them {\em admissible}) of the form
\begin{equation} \label{gens} v_{0}^{\epsilon}v_{1}^{m}v_{2}^{n}\sigma^{2p},
\end{equation} 
where $m=\{0,1,2,...\}$, $n\in\Z$, $\epsilon=0$ if $m>0$ and $p$ is even or
$p$ is divisible by $4$ and $\epsilon=1$ otherwise. The only relation on the
$\Z[a]/(2a)$-module generators (\ref{gens}) is that those with $\epsilon=1$
are annihilated by $a$. The dimension of each element (\ref{gens}) is
determined by setting the dimension of $v_{i}$ equal to $(2^i-1)(1+\alpha)$
($i=0,1,2$). The dimension of $\sigma$ is $\alpha-1$. The notation for the
generators (\ref{gens}) comes from a spectral sequence used to calculate
$E\mathbb{R}(2)_{\star}$. This notation suggests that the multiplicative
structure on $E\mathbb{R}(2)_{\star}$ could be computed simply by
multiplying the monomials (\ref{gens}). This multiplicative structure is in
effect right, if we put $v_{0}^{2}=2v_{0}$ (\cite{hk}).

\vspace{3mm}
To make (\ref{gens}) seem a little less exotic, it may be helpful to note
that the coefficients $KR_{\star}$ of Atiyah's real $KR$-theory have an
analogous
description: the $\Z[a]/(2a)$-generators are
$$v_{0}^{\epsilon}v_{1}^{n}\sigma^{2p}$$
where $n\in\Z$, and $\epsilon=0$ if $p$ is even and $\epsilon=1$ if $p$ is
odd. One has the relation $v_0 a=0$ and $v_{0}^{2}=2v_{0}$.

\vspace{3mm}
In this paper, we have considered real-oriented elliptic cohomolgoy $EO(2)$,
which can be thought of as the ``fixed points'' of elliptic cohomology with
respect to the formal inverse. However, there is more to the story. Hopkins
and Miller \cite{Hm}, see also \cite{tmf} 
have constructed an even much more elaborate
generalized cohomology theory $tmf$ (which stands for {\em topological
modular forms}). Roughly speaking, to understand $tmf$, one must study the
coordinate changes of the Weierstrass curve. These changes may be encoded in
an affine algebraic groupoid, the ring of whose coefficients turns out to
coincide with the $0$-dimensional part of the coefficients of a cosimplicial
elliptic spectrum $tmf_*$; its realization is $tmf$. The construction of the
cosimplicial 
elliptic spectrum $tmf_*$
(in particular making that the cosimplicial structure strict, not just up to
homotopy) is the deep part of the theory, and is the subject of the work of
\cite{Hm}, much of which is still unpublished. The theory $tmf$ has the
striking property that it is 
$MO\langle 8\rangle$-orientable,
i.e. a manifold is $tmf$-orientable if it is $Spin$ and has $p_1/2=0$. This
was proved by Ando, Hopkins and Rezk in \cite{ahr}, and a part of their
theory was also used in the present paper in Section \ref{elleo} above.

\end{appendix}




\begin{thebibliography}{99} 
\bibitem{dmw}
E.~Diaconescu, G.~Moore and E.~Witten,
{\it $E_8$ gauge theory, and a derivation of K-Theory from M-Theory}, Adv.
Theor. Math. Phys. {\bf 6} (2003) 1031, [{\tt arXiv:hep-th/0005090}].

\bibitem{Pol}
J. Polchinski,
{\it Dirichlet-branes and Ramond-Ramond charges},
Phys. Rev. Lett. {\bf 75} (1995) 4724,
[{\tt arXiv:hep-th/9510017}].

\bibitem{MM}
R.~Minasian and G.~Moore,
{\it K-theory and Ramond-Ramond charge},
J. High Energy Phys. {\bf 11} (1997) 002,
[{\tt arXiv:hep-th/9710230}].

\bibitem{Wi1}
E.~Witten,
{\it D-Branes and K-Theory},
J. High Energy Phys. {\bf 12} (1998) 019, 
[{\tt arXiv:hep-th/9810188}].

\bibitem{Ho}
P.~Ho\v{r}ava, {\it Type $IIA$ D-branes, K-theory, and matrix theory},
Adv.~Theor.~Math.~Phys.~{\bf 2} (1999) 1373, [{\tt arXiv:hep-th/9812135}].

\bibitem{MW}
G.~Moore and E.~Witten,
{\it Self duality, Ramond-Ramond fields, and K-theory},
J. High Energy Phys. {\bf 05} (2000) 032,
[{\tt arXiv:hep-th/9912279}].

\bibitem{FH}
D.~S.~Freed and M.~J.~Hopkins, {\it On Ramond-Ramond fields and K-theory},
J. High Energy Phys. {\bf 05} (2000) 044, [{\tt arXiv:hep-th/0002027}]. 

\bibitem{dk}
P.~Donovan and M.~Karoubi,
{\it Graded Brauer groups and $K$-theory with local coefficients},
Inst.~Hautes~\'Etudes~Sci.~Publ.~Math. {\bf 38} (1970) 5.

\bibitem{Ros}
J.~Rosenberg, {\it Continuous trace $C^*$-algebras from
the bundle theoretic point of view},
J. Aust. Math. Soc. {\bf A47} (1989) 368.

\bibitem{FW}
D.~S.~Freed and E.~Witten, {\it Anomalies in string theory with D-Branes},
Asian J. Math. {\bf3} (1999) 819, [{\tt arXiv:hep-th/9907189}].

\bibitem{Ka} 
A.~Kapustin, {\it D-branes in a topologically nontrivial B-field}, Adv.
Theor. Math.~Phys.~{\bf 4} (2000) 127, [{\tt arXiv:hep-th/9909089}].

\bibitem{BM}
P. Bouwknegt and V. Mathai,
{\it D-branes, B-fields and twisted K-theory},
J. High Energy Phys. {\bf 03} (2000) 007,
[{\tt arXiv:hep-th/0002023}].

\bibitem{guk}
S.~Gukov,
{\it K-theory, reality, and orientifolds},
Commun. Math. Phys. {\bf 210} (2000) 621,
[{\tt arXiv:hep-th/9901042}].

\bibitem{ho}
K.~ Hori,
{\it D-branes, T-duality, and index theory},
Adv. Theor. Math. Phys. {\bf 3} (1999) 281,
[{\tt arXiv:hep-th/9902102}].

\bibitem{sz}
K.~Olsen and R.~Szabo,
{\it Constructing D-branes from K-Theory},
Adv. Theor. Math. Phys. {\bf 3} (1999) 889,
[{\tt arXiv:hep-th/9907140}].

\bibitem{m1}
E.~Witten, {\it String theory dynamics in various dimensions}, Nucl.~Phys.
{\bf B443} (1995) 85, [{\tt arXiv:hep-th/9503124}].

\bibitem{m2}
P.~K.~Townsend, {\it Four lectures on M-theory}, in Trieste summer school 
in High Energy Physics and Cosmology, 1996, 
[{\tt arXiv:hep-th/9612121}].

\bibitem{m3}
M.~J.~Duff,
{\it M-theory (the theory formerly known as strings)},
Int. J. Mod. Phys. {\bf A11} (1996) 5623,
[{\tt arXiv:hep-th/9608117}].

\bibitem{flux}
E.~Witten,
{\it On flux quantization in M-theory and the effective action}, J. Geom.
Phys. {\bf 22} (1997) 1, [{\tt arXiv:hep-th/9609122}].

\bibitem{duality}
E.~Witten,
{\it Duality relations among topological effects in string theory}, J. High
Energy Phys. {\bf 0005} (2000) 031, [{\tt arXiv:hep-th/9912086}].

\bibitem{hs}
M.J. Hopkins, I.M. Singer,
{\it Quadratic functions in geometry, topology,and M-theory}, [{\tt
arXiv:math.AT/0211216}].

\bibitem{dfm}
 E.~Diaconescu, D.~Freed and G.~Moore,
{\it The M-theory 3-form and E8 gauge theory},
[{\tt arXiv:hep-th/0312069}].

\bibitem{ms}
G.~Moore and N.~Saulina,
{\it T-duality, and the K-theoretic partition function of Type IIA
superstring theory}, Nucl. Phys. {\bf B670} (2003) 27, [{\tt
arXiv:hep-th/0206092}].

\bibitem{hm}
V.~Mathai and H.~Sati,
{\it Some relations between twisted K-theory and $E\sb8$ gauge theory}, J.
High Energy Phys. {\bf 03} (2004) 016, [{\tt arXiv:hep-th/0312033}].

\bibitem{doug}
M.~R.~Douglas,
{\it D-branes, categories and $N=1$ supersymmetry},
J. Math. Phys. {\bf 42} (2001) 2818.
[{\tt arXiv:hep-th/0011017}].

\bibitem{katz}
A.~Caldararu, S.~Katz and E.~Sharpe,
{\it D-branes, B fields, and Ext groups},
Adv. Theor. Math. Phys. {\bf 7} (2003) 381,
[{\tt arXiv:hep-th/0302099}].

\bibitem{asp}
P.~Aspinwall and A.~Lawrence, 
{\it Derived categories and zero-brane stability}, 
J. High Energy Phys. {\bf 0108} (2001) 004,
[{\tt arXiv:hep-th/0104147}].

\bibitem{dist}
I~Brunner and J.~Distler,
{\it Torsion D-branes in nongeometrical phases},
Adv. Theor. Math. Phys. {\bf 5} (2002) 265,
[{\tt arXiv:hep-th/0102018}].

\bibitem{lot}
J.~Lott,
{\it $R/Z$ index theory},
Comm. Anal. Geom. {\bf 2} (1994), no. 2, 279.

\bibitem{kil}
T.~Killingback,
{\it Global anomalies, string theory and spacetime topology}, Class. Quant.
Grav. {\bf 5} (1988), 1169.

\bibitem{st} 
S.~Stolz and P.~Teichner,
{\it What is an elliptic object?},
to appear in the Proceedings of Graeme Segal's Birthday Conference, 
Oxford University Press, www.math.ucsd.edu/~teichner/Preprints/Oxford.pdf . 

\bibitem{bp}
E.~H.~Brown and F.~P.~Peterson,
{\it A spectrum
whose $\Z/p$ cohomology is the algebra of reduced $p$-th powers}, Topology
{\bf 5} (1966) 149.

\bibitem{mil} 
J.~Milnor,
{\it The Steenrod algebra and its dual},
Ann.~of~Math. (2) {\bf 67} (1958) 150.

\bibitem{rav} 
D.~Ravenel,
{\it Complex cobordism and stable homotopy groups of spheres}, 
Academic Press, Inc., Orlando, FL, 1986.

\bibitem{BP2}
E.~Brown, Jr. and F.~Peterson,
{\it Relations among characteristic classes. I },
Topology {\bf 3} (1964) suppl. 1, 39.

\bibitem{Hm}
M.~J.~Hopkins, H.~R.~Miller, M.~Mahowald and P.~Goerss, {\it in
preparation}.

\bibitem{tmf}
M.~Ando, H.~J.~Hopkins and N.~P.~Strickland, 
{\it Elliptic spectra, the Witten genus and the theorem of the cube}, 
Invent. math. {\bf 146} (2001) 595.

\bibitem{lt}
J.~Lubin and J.~Tate,
{\it  Formal moduli for one-parameter formal Lie groups},
Bull. Soc. Math. France {\bf 94} (1966) 49.

\bibitem{hk}  
P.~Hu and I.~Kriz,
{\it Real-oriented homotopy theory
and an analogue of the Adams-Novikov spectral sequence}, Topology {\bf 40}
(2001) 317.

\bibitem{ahr}
M.~Ando, M.~J.~Hopkins, C.~Rezk, 
{\it Orientability with respect to $tmf$},
in preparation.

\bibitem{rezk}
C.~Rezk, 
{\it The units in a ring spectrum and the logarithm}, preprint, 2004.

\bibitem{kuhn}
N.~Kuhn, {\em Morava $K$-theories and infinite loop spaces}, Algebraic
topology (Arcata, CA, 1986), Lecture Notes in Mathematics, Vol. {\bf 1370},
Springer Verlag, Berlin, 1989, pp. 243-257.

\bibitem{ando}
M.~Ando, {\em Isogenies of formal group laws and power operations in the
cohomology theories $E_n$}, Duke Math. J. {\bf 79} (1995), 423.

\bibitem{mqr} J.~P.~May, {\em $E_{\infty}$-ring spaces and $E_{\infty}$-ring
spectra}, with contributions by F.~Quinn, N.~Ray and J.~Torenhave, Lecture
Notes in Mathematics, Vol. {\bf 577}, Springer Verlag, Berlin, 1977.

\bibitem{segal}
G.~Segal,
{\it Elliptic cohomology},
S\'{e}minaire Bourbaki, Vol. 1987/88, Ast\'{e}risque {\bf 161-162} (1988),
Exp. No. {\bf 695} (1989) 187.

\bibitem{mor}
J.~Morava,
{\it Forms of $K$-theory}, Math.~Z. {\bf 201} (1989) 401.

\bibitem{bdr}
N.~Baas, B.~A.~Dundas and J.~Rognes,
{\it Two-vector bundles and
forms of elliptic cohomology}, 
to appear in the Proceedings of Graeme Segal's Birthday Conference, 
Oxford University Press, [{\tt arXiv:math.AT/0306027}].

\bibitem{prog}
I.~Kriz and H.~Sati,
{\it work in progress}.

\bibitem{dtwist}
C.~L.~Douglas,  
{\it On the twisted $K$-homology of simple Lie
groups}, [{\tt arXiv:math.AT/0402082}].

\bibitem{hkdb}
P.~Hu, I.~Kriz, 
{\it Closed and open conformal field theories and their anomalies}, to
appear in Commun. Math. Phys., [{\tt arXiv:hep-th/0401061}].

\bibitem{Lech}
K. Lechner, P. A. Marchetti and M. Tonin,
{\it Anomaly free effective action for the elementary M5-brane}, Phys. Lett.
{\bf B524} (2002) 199, [{\tt arXiv:hep-th/0107061}].


\bibitem{jerome}
J.~Gauntlett, 
{\it Intersecting branes},
lectures at APCTP winter school on Dualities of Gauge and String Theories,
Sokcho, Korea, 1997,  [{\tt arXiv:hep-th/9705011}].

\bibitem{om}
R.~Gopakumar, S.~Minwalla, N.~Seiberg and A.~Strominger,
{\it OM theory in diverse dimensions},
J. High energy Phys. {\bf 08} (2000) 008,
[{\tt arXiv:hep-th/0006062}].

\bibitem{hkconf}
P.~Hu and I.~Kriz,
{\it  Conformal field theory and elliptic cohomology},
to appear in Advances in Mathematics.

\bibitem{sen}
A.~Sen,
{\it T-duality of p-branes},
Mod.~Phys.~Lett. {\bf A11} (1996) 827,
[{\tt arXiv:hep-th/9512203}].

\bibitem{ganor}
O.~Ganor, S.~Ramgoolam and W.~Taylor IV,
{\it Branes, fluxes and duality in M(atrix)-theory}, Nucl.~Phys. {\bf B492}
(1997) 191, [{\tt arXiv:hep-th/9611202}].

\bibitem{rud}
Y.~Rudyak,
{\it On Thom spectra, orientability, and cobordism}, Springer-Verlag,
Berlin, 1998.

\bibitem{bous}
A.~K. Bousfield, {\em The localization of spectra with respect to homology},
Topology 19 (1979), 257-281

\bibitem{hopm}
M.~J.~Hopkins and M.~Mahowald, 
{\it From elliptic curves to homotopy theory},
preprint, 1998,
http://hopf.math.purdue.edu/cgi-bin/generate?/Hopkins-Mahowald/eo2homotopy .


\end{thebibliography}
\end{document}